\documentclass{aa}

\usepackage{graphicx,amssymb,natbib}
\usepackage{txfonts}
\usepackage{mathrsfs}

\usepackage{color}

\begin{document}

\def\A{\,\AA\ }
\def\AF{\,\AA}
\newcommand{\etal}{et~al. }
\newcommand{\ca}{\ion{Ca}{ii} 8542\A}
\newcommand{\fca}{\ion{Ca}{ii} 8542~\AA}
\newcommand{\cmv}{\mbox{{\rm\thinspace cm$^{-3}$}}}
\newcommand{\cmmf}{\mbox{{\rm\thinspace cm$^{-5}$}} }
\newcommand{\D}{\displaystyle}
\newcommand{\eq}[1]{Eq.\,(\ref{#1})}
\newcommand{\fig}[1]{Fig.~\ref{#1}}
\newcommand{\km}{\mbox{{\rm\thinspace km}}}
\newcommand{\kms}{\mbox{{\rm\thinspace km\thinspace s$^{-1}$}}}
\newcommand{\ms}{\mbox{{\rm\thinspace m\thinspace s$^{-1}$}}}
\newcommand{\K}{\mbox{{\thinspace\rm K}}}
\newcommand{\tab}[1]{Table~\ref{#1}}

\newcommand{\ben}{\begin{enumerate}}
\newcommand{\een}{\end{enumerate}}
\newcommand{\bfig}{\begin{figure}}
\newcommand{\efig}{\end{figure}}
\newcommand{\beq}{\begin{equation}}
\newcommand{\eeq}{\end{equation}}
\newcommand{\mbf}{\mathbf}
\newcommand{\vare}{\varepsilon}
\newcommand{\mcal}{\mathcal}
\newcommand{\ep}{\epsilon}
\newcommand{\cs}{\mathcal{S}}
\newcommand{\csv}{\mathcal{S}_{\mathcal{V}}}
\newcommand{\cv}{\mathcal{V}}
\renewcommand{\thefootnote}{\dag}
\newcommand{\ha}{${\rm H\alpha}$}

\title{A persistent quiet-Sun small-scale tornado}
\subtitle{II. Oscillations}

\author{K.\,Tziotziou\inst{1}
        \and G.\,Tsiropoula\inst{1} \and I.\,Kontogiannis\inst{2}}

\offprints{K.\,Tziotziou,\\
\email{kostas@noa.gr}}

\institute{Institute for Astronomy, Astrophysics, Space Applications and Remote Sensing,
National Observatory of Athens, GR-15236 Penteli, Greece \and Leibniz-Institut f\"{u}r Astrophysik Potsdam (AIP), An der Sternwarte 16, 14482 Potsdam, Germany}

\date{Received  / Accepted }

\titlerunning{Oscillations in a persistent quiet-Sun tornado}

\authorrunning{Tziotziou \etal}

\abstract{Recently, the appearance, characteristics, and dynamics of a persistent 1.7~h vortex flow, resembling a small-scale tornado, have been investigated with observations both from the ground and from space in a quiet-Sun region in several lines and channels and for the first time in the \ha\ line centre. The vortex flow showed significant substructure in the form of several intermittent chromospheric swirls.}
{We investigate the oscillatory behaviour of various physical parameters in the vortex area in an attempt to better understand the physics of the reported vortex flow. This is the first analysis of this extent.}
{We used the same data set of high spatial and temporal resolution CRisp Imaging SpectroPolarimeter (CRISP) observations in several wavelengths along the \ha\ and \ca line profiles, as well as Doppler velocities and full-width at half-maximum (FWHM) derived from the \ha\ line profiles. The spectral analysis of oscillations is based on a two-dimensional wavelet analysis performed within the vortex flow area and in a quiet-Sun region (used for comparison), as well as along line and circular slices.}{The vortex flow shows significant oscillatory power in the range of 3 to 5~min, peaking around 4~min. This power behaves differently than the reference quiet-Sun region. The derived oscillations reflect the cumulative action of different components such as swaying motions, rotation, and waves. The derived periods for swaying motions are in the range of 200--220~s, and the rotation periods are $\sim$270~s for \ha\ and $\sim$215~s for \fca. Periods increase with atmospheric height and seem to decrease with radial distance from the vortex centre, suggesting a deviation from a rigid rotation. The behaviour of power within the vortex flow as a function of period and height implies the existence of evanescent waves. Moreover, considerable power is obtained even for periods as long as 10~min, not only at photospheric but also at chromospheric heights, while the formation of vortexes is related to turbulent convection or to twisting motions exercised in the magnetic field concentrations. These imply that different types of waves may be excited, such as magnetoacoustic (e.g. kink) or Alfv\'{e}n waves.}
{The vortex flow seems to be dominated by two motions: a transverse (swaying) motion, and a rotational motion. The obtained oscillations point to the propagation of waves within it. Nearby fibril-like flows could play an important role in the rotational modulation of the vortex flow. There also exists indirect evidence that the structure is magnetically supported, and one of the swirls, close to its centre, seems to be acting as a ``central engine'' to the vortex flow.}


\maketitle

\section{Introduction}
\label{intro}

The highly turbulent nature of convective motions creates many complicated patterns that are visible in various spatial and temporal scales in the solar atmosphere. Observations of swirling motions that became possible thanks to modern instrumentation with high spatial, temporal, and spectral resolution have attracted increased interest in the past years.
These motions, also known as vortex motions, have been predicted by theory \citep{Sten:1975} and found in numerical simulations \citep[e.g.][]{Stein:2000b, Moll:2011b, Moll:2011, Kitia:2012, Wede:2012}.  According to realistic 3D magnetohydrodynamic (MHD) numerical simulations, ubiquitous small-scale vortex tubes are generated by turbulent flows  in the solar subsurface layers that are concentrated at the vertices of intergranular lanes as a direct consequence of the conservation of angular momentum. Their interaction with magnetic fields leads to various rotating structures that extend from the photosphere to the chromosphere and corona.

Observations have shown that vortex flows are found almost everywhere in the solar atmosphere. At the photospheric level, they manifest as small- and large-scale vortex-like motions \citep[e.g.][]{Brandt:1988, Bonet:2008, Bonet:2010, Attie:2009} that are found in granular and supergranular junctions. \citet{Bonet:2008} traced the motions of magnetic bright points (BPs) in G-band images obtained with the CRisp Imaging SpectroPolarimeter \citep[CRISP;][]{Scharmer:2008} instrument mounted at the Swedish 1m Solar Telescope \citep[SST;][]{Scharmer:2003a} and showed that some of them follow logarithmic spiral trajectories. Bonet and collaborators concluded that these trajectories represent convection-driven vortex flows created at the downdrafts where the hot plasma that rises from the solar interior returns to the solar surface after cooling down. \citet{Wede:2009} analysed observations of image sequences obtained with CRISP and showed structures with clear rotational motions in the core of the \ca line that maps the chromosphere. These events are referred to as ``chromospheric swirls''. Responses to such swirls have also been identified above the chromosphere in the transition region and up to the low corona through observations in several UV and EUV channels of the Atmospheric Imaging Assembly \citep[AIA;][]{Lemen:2012} on the Solar Dynamics Observatory \citep[SDO;][]{Pesnell:2012} space mission \citep{Wede:2012}. Small-scale vortices have recently also been observed by \cite{Park:2016} simultaneously in the wings of the \ha\ line and in the cores of the \ha\ and \ca lines, as well as in the strong ultraviolet (UV) \ion{Mg}{ii}~k and Mg\,{\sc ii} subordinate lines observed by the space-based Interface Region Imaging Spectrograph \citep[IRIS;][]{depont:2014}.
Recently, \cite{tzio:2018} (hereafter Paper I) used observations with high spatial and temporal resolution obtained with CRISP in several wavelengths along the \ha\ and \ca line profiles and simultaneous SDO/AIA observations in several UV and EUV channels and Helioseismic and Magnetic Imager \citep[HMI;][]{Scher:2012} magnetograms to investigate the appearance, characteristics, and dynamics of a persistent vortex flow. The flow was best seen in the \ha\ line centre and close line-centre wavelengths, and its most important characteristic was its long duration (at least 1.7~h) and its large radius ($\sim$3\arcsec). Its rotation period was roughly estimated to be in the range of 200 to 300~s. For an overview of the vortex flow characteristics presented in Paper~I, see Sect.~\ref{over}.

The vortex tubes can capture and amplify the magnetic field, penetrate chromospheric and coronal layers, affect the thermodynamic properties of their environment, and trigger various dynamical effects, such as spontaneous quasi-periodic (with periods 2-5~min) flow eruptions \citep{Kitia:2013} that generate shocks in the atmosphere. \citet{Bonet:2010} used observations obtained by the Imaging Magnetograph eXperiment \citep[IMaX;][]{Mart:2011} on board the Sunrise balloon \citep{Barth:2011} to detect proper motions of individual vortices and reported rotation periods of 35~min. \citet{Moll:2011b} have shown in near-surface solar convection simulations that the peak of the swirling period increases with height and attains a peak value of $\sim$220~s at a height $z$=90~km. Periodic motions at the vortex tube footpoints excite a wide variety of MHD wave modes and oscillations. For example, \citet{Fedun:2011c} showed that the implementation of high-frequency twisting of an open flux tube by photospheric vortices (period of 30~s) generates a broadband frequency range of incompressible MHD waves (i.e. torsional Alfv\'{e}n and transverse/kink waves). \cite{Fedun:2011b} applied torsional drivers with periods in the range of 2 to 6~min and demonstrated how magnetic flux tubes can act as a spatial frequency filter transporting selective periods of torsional Alfv\'{e}n waves.

Rotational motions are very important because they can play a key role in energy, mass, and momentum transfer between the subsurface and the upper layers of the Sun; they may also provide an alternative mechanism for supplying energy to the corona. Their complicated dynamics can generate various types of oscillations that can be identified in physical parameters, such as intensity and velocity, that are related to them. Detections of oscillatory motions in vortexes from imaging observations can provide an important diagnostic tool for the propagation of waves and the exploration of their role in the energy flux that is carried upwards from the solar interior.

In this paper, we expand the analysis of the vortex flow presented in Paper~I. Our aim is to investigate the oscillatory behaviour of various physical parameters in the vortex area and investigate their behaviour in several atmospheric heights in an attempt to better understand the physics of the reported vortex flow and answer related questions stemming from our previous work. To our knowledge, such an analysis has not been performed so far. The presence, signatures, and propagation of waves within the vortex area will be investigated in a follow-up paper.

\begin{figure*}
\centerline{\includegraphics[width=0.7\textwidth]{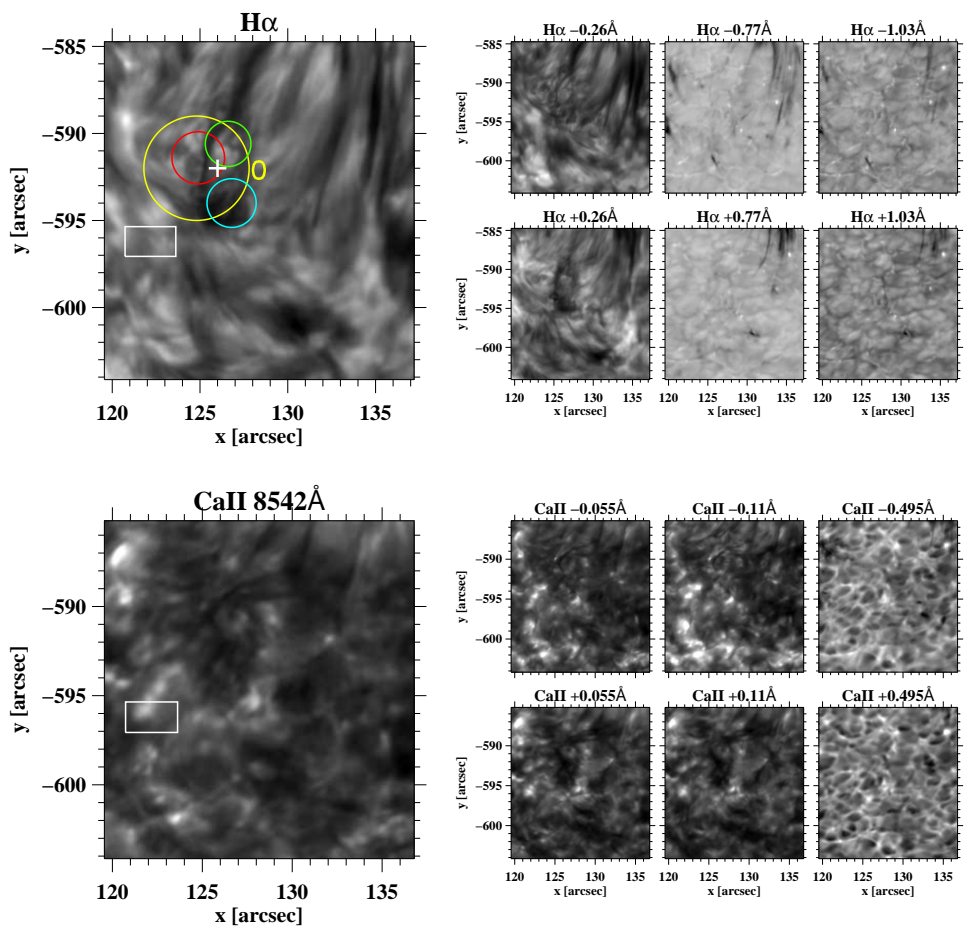}}
\caption{Snapshot of the ROI in all acquired wavelengths along the \ha\ (top row) and the \ion{Ca}{ii} 8542$\,${\AA} (bottom row) line profiles, with black indicating structures in absorption. The overplotted yellow circle in the line-centre \ha\ panel indicates the location of the analysed conspicuous vortex flow, and smaller red, green, and cyan circles denote the approximate location of smaller swirls (substructure) that we described in Sect.~\ref{over} and extensively discussed in Paper I. The white rectangle shows the reference QSR (see text), and the white cross indicates the location of the wavelet spectra presented in \fig{fig2}. The yellow 0 indicates the start 0\degr\ of the clockwise angles used for the x-axis of \fig{fig4}. } \label{fig1}
\end{figure*}

\section{Observations and methodology}
\label{obsmeth}

\subsection{Observations}
\label{obs}

For our analysis we used the same data sets as in Paper I, comprising two high-cadence (i.e. 4 seconds) high-spatial resolution, multi-wavelength imaging spectroscopy time series of a quiet-Sun region, obtained on June 7, 2014, between 07:32 UT -- 08:21 UT and 08:28 UT -- 09:16 UT with the CRISP at the SST. The spectral images are sampled at seven wavelength positions along two line profiles: the \ha\ 6562.81$\,${\AA} line profile at the line centre, \ha$\pm0.26\,${\AA}, \ha$\pm0.77\,${\AA}, and \ha$\pm1.03\,${\AA} and the \ion{Ca}{ii} 8542$\,${\AA} line profile at the line centre, \ion{Ca}{ii}$\pm$0.055$\,${\AA}, \ion{Ca}{ii}$\pm$0.11$\,${\AA}, and \ion{Ca}{ii}$\pm$0.495$\,${\AA}. The scale resolution is 0.059$\arcsec$ and 0.0576$\arcsec$ per pixel for the \ha\ and \ion{Ca}{ii} 8542$\,${\AA} lines, respectively, while there exists a small temporal offset of $\sim$2~s between the \ion{Ca}{ii} and \ha\ time series. No UV, EUV, and magnetogram data obtained with the AIA and HMI instruments on board the SDO were used for this analysis because their spatial and temporal resolution is lower than that of the CRISP observations.

The region of interest (ROI), containing the analysed vortex flow (see \fig{fig1} in this paper and Paper I), is located in the south-west solar hemisphere, almost at the centre of the acquired 60$\arcsec\times\,$60$\arcsec$ CRISP FOV and at the centre of a supergranule, and covers an area of 17.3$\arcsec\times\,$19$\arcsec$.
The ROI centre lies at a latitude $\sim$38\degr\ south of the solar equator and a longitude of $\sim$10\degr\ west of the central meridian.
Figure~\ref{fig1} shows snapshots of the ROI in the observed seven wavelengths along the \ha\ and \ion{Ca}{ii} line profiles, with the location of the analysed vortex flow clearly marked by a yellow circle. The approximate centre of the vortex flow is at (x,$\,$y) $=$ (124.8$\arcsec$,$\,-$592$\arcsec$), and it has a radius of $\sim$3\arcsec.
Figure~\ref{fig1} further indicates the quiet-Sun region (QSR) that we used as a reference region for the analysis in this paper. Because the whole ROI is extremely dynamic, a small region that is void of large fibril-like structures or small vortex flows was carefully selected.

We refer to Paper I for further details concerning the observations, the processing of the datacubes, and their co-alignment.

\subsection{Methodology}
\label{meth}

In addition to the intensities at different wavelengths in the two spectral lines, we here further derive Doppler velocities and the full-width at half-maximum (FWHM) of the \ha\ profiles, and perform a spectral analysis to investigate their oscillatory behaviour. Below we briefly describe the methods we used to derive these physical quantities and determine their oscillatory behaviour.

\paragraph{\ha\ Doppler velocities}

\ha\ Doppler velocities at each pixel were derived from the shift of the minimum of the cubic-spline interpolated \ha\ profiles with respect to the minimum of the spatial profile average over the whole ROI for each image of the time series. No Doppler velocities and FWHM could be derived from the \ca line profiles because the wavelength sampling coverage is only regular close to the line centre, and moreover, there is no wavelength sampling further than $\pm$0.495$\,${\AA}; thus, a proper cubic-spline interpolation of the \ion{Ca}{ii} line profile would not give reliable results. We refer to Paper~I for further details concerning the derivation of \ha\ Doppler velocities and the zero-reference wavelength. We also recall that as discussed in Paper I, the acquired Doppler velocities are only a qualitative and not a quantitative representation of the actual LOS velocity as the wavelength coverage is quite coarse, the shape of the H$\alpha$ line profile also reflects significant temperature and opacity contributions, and projection effects also exist because the ROI is located far away from the disc centre.

\paragraph{FWHM of the \ha\ profiles}
The observed \ha\ line profiles do not reach the local continua. This prevents us from measuring the profile width at a fixed fractional depth from the continuum. We therefore used the
method of \cite{Cauzzi:2009} to derive the \ha\ FWHM, applied to the cubic-spline interpolated \ha\ profiles.
For each such temporal and spatial \ha\ line profile, the wing intensities at a specified wavelength separation $\Delta\lambda$ from the profile
minimum were averaged, and the FWHM was defined as the wavelength separation of the profile samplings at half the intensity range between the profile
minimum and this wing average \citep[see Fig.~2 of][]{Cauzzi:2009}. This wavelength separation $\Delta\lambda$ is rather arbitrary and fixed at 0.8 \AA\ because the profiles extend up to $\pm1.03$ \AA, and it therefore allows deriving the FWHM even for profiles that have a Doppler velocity amplitude of up to 8\kms\ (see Fig.~8 of Paper~I).

\begin{figure*}
\centerline{\includegraphics[width=0.3\textwidth]{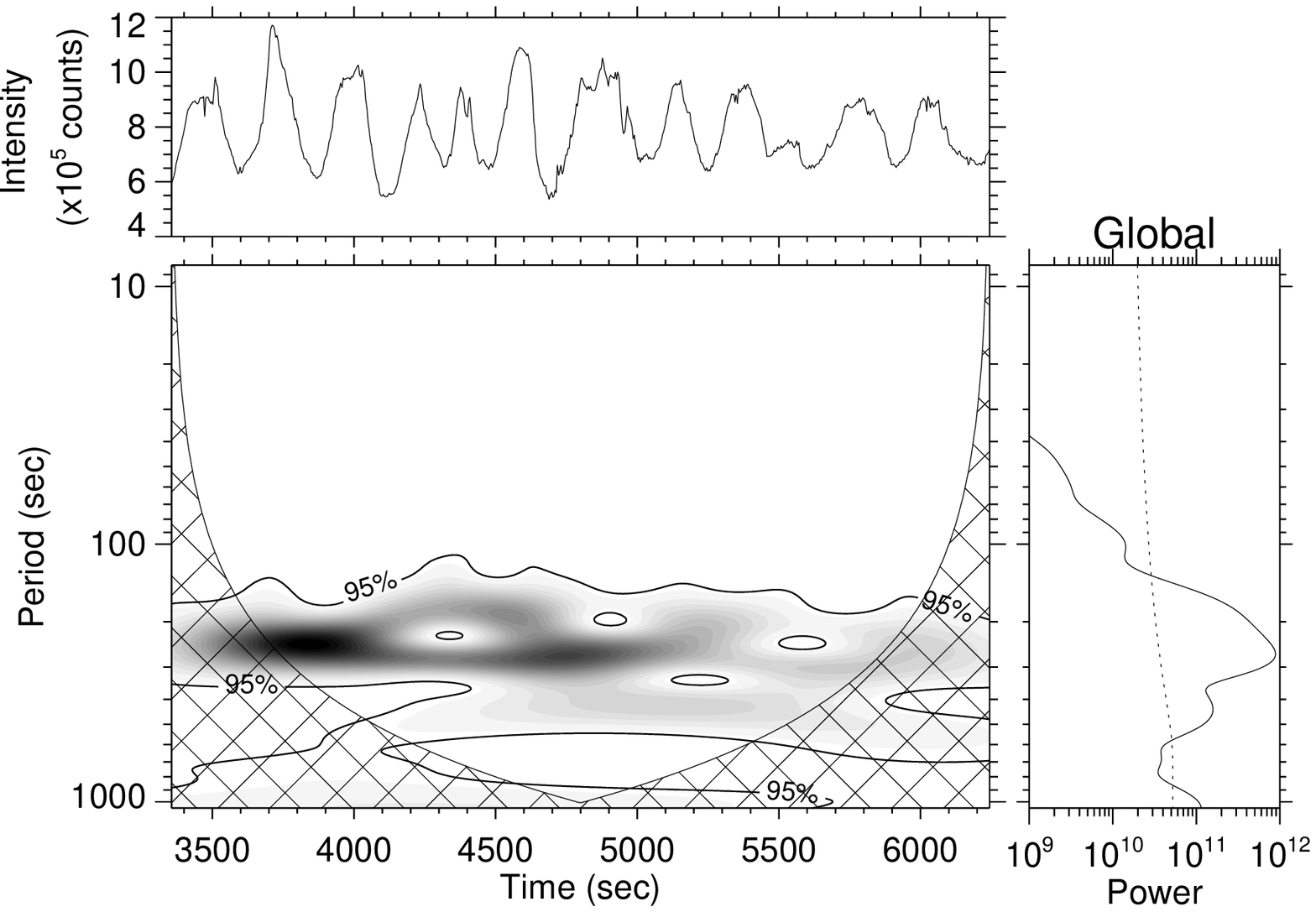}\hspace{0.8cm}\includegraphics[width=0.3\textwidth]{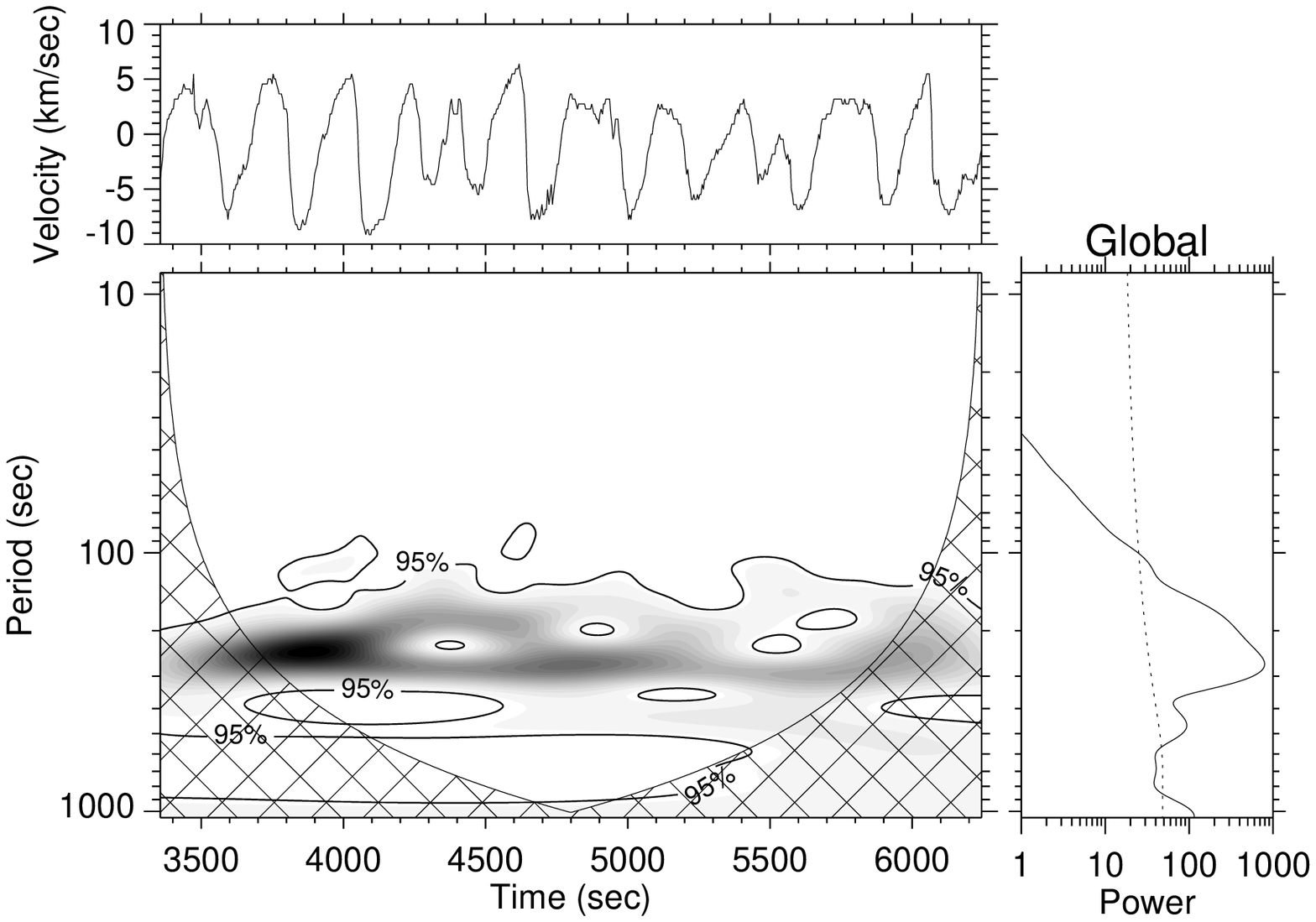}
\hspace{0.8cm}\includegraphics[width=0.3\textwidth]{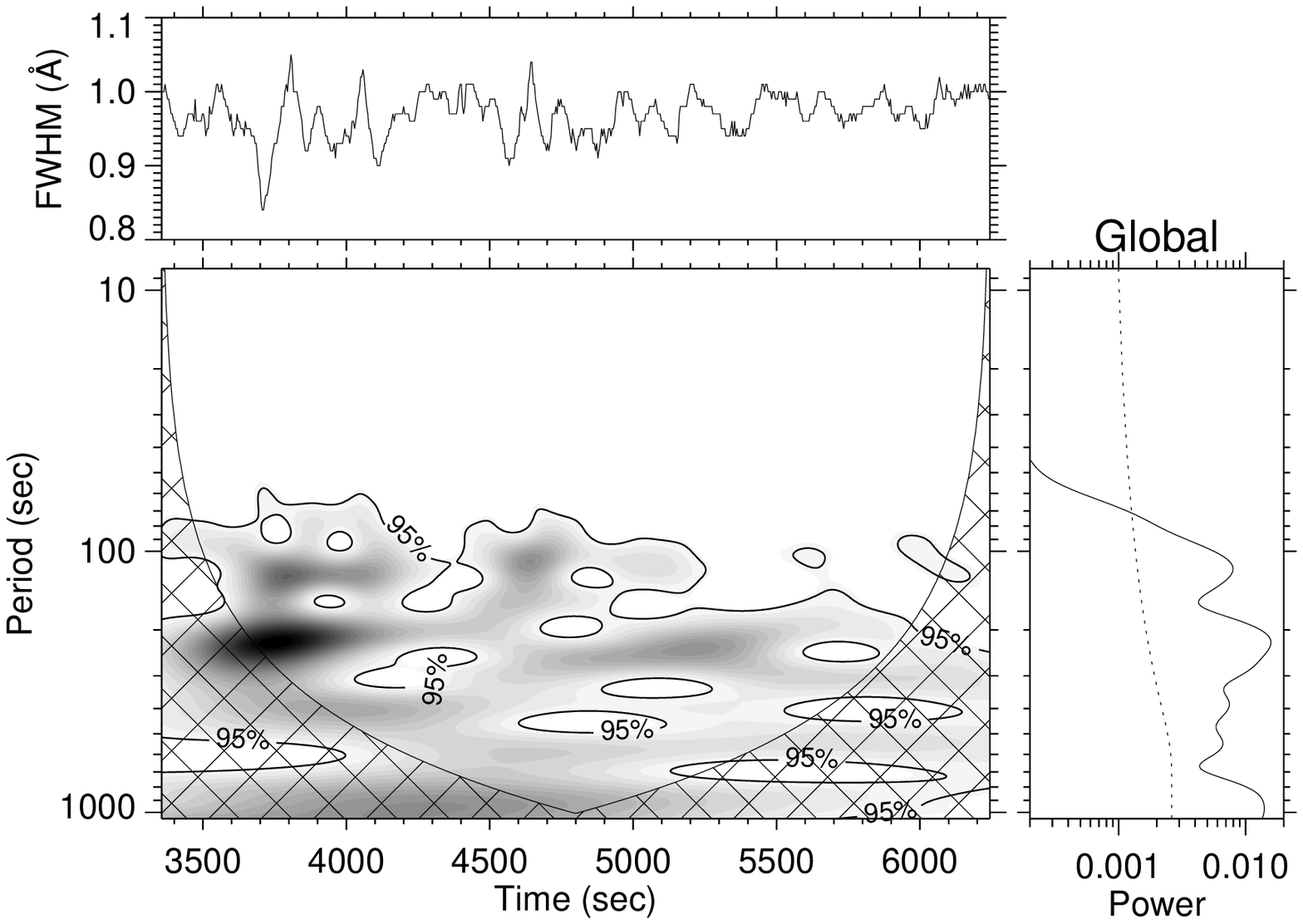}}
\caption{Indicative wavelet analysis figures for the \ha$-$0.26$\,${\AA} intensity, \ha\ Doppler velocity, and \ha\ FWHM variations (left, middle, and right figures, respectively) during the second observing interval (08:28 UT -- 09:16 UT) at location (x,$\,$y) $=$ (126$\arcsec$,$\,-$592$\arcsec$), indicated by a white cross in \fig{fig1}. Time is in seconds starting at 07:32 UT. In each figure the top panels show the time variations of the analysed quantity (intensity, Doppler velocity, or FWHM), and the bottom left panels show the time-period calculated power spectra for the variations shown in the top panels. Filled contours correspond to the derived spectra; black represents high power values, and solid contours represent the 95\% significance level. Cross-hatched regions indicate regions where edge effects of our finite time-series may become important, therefore periods within these regions should be treated with caution. The right panels in each figure show the global power spectrum, that is, the average of the wavelet power spectrum over time with the dashed line indicating the respective global significance level of 95\%.}\label{fig2}
\end{figure*}

\paragraph{Spectral analysis}
To study the oscillatory behaviour of the different physical quantities, we used a wavelet analysis \citep{Torr:1998}, which is a suitable technique for determining any present periodic signal and its variation with time. We used the Morlet wavelet function (sinusoid modulated by a Gaussian) because it is appropriate and most commonly used for harmonic-type
oscillations. We refer to \cite{Tzio:2004} for further details about the wavelet method and the determination of the significance levels. \fig{fig2} shows indicative wavelet power spectra for \ha\ intensity, velocity, and FWHM variations at a certain location within the vortex area (marked in \fig{fig1}) that we discuss in Sect.~\ref{osci} in more detail. In addition to the derived power spectra as a function of time and period, the global power spectrum, that is, the average of the wavelet power spectrum over time, was also derived (see the respective panels of \fig{fig2}) and used to determine the dominant oscillating periods (maxima of the global power spectrum). Any discussion hereafter about power spectra refers to these derived global power spectra unless specified otherwise.

\section{Overview of the vortex flow characteristics}
\label{over}

The analysis in Paper I revealed the existence of a funnel-like vortex structure that expands with height, extends from the photosphere to the low corona, and rotates clockwise rigidly or quasi-rigidly. No clear evidence, however, was found for a magnetically driven flow because no associated magnetic BPs were observed in the photosphere. The striking characteristics of this vortex flow, which has been revealed for the first time in \ha\ observations at the line centre and near the line-centre wavelengths, are its long duration ($\geq$1.7 h), its large radius ($\sim$3\arcsec), and the existence of significant substructure within it. It resembles a small-scale quiet-Sun tornado, in contrast to previously reported short-lived swirls and in analogy to persistent giant tornadoes. The observed substructure is visible as several individual intermittent chromospheric swirls that show typical sizes and durations and a wide range of morphological patterns mostly associated with upwards velocities up to 8\kms\ and a mean value of $\sim$3\kms. The different dynamics of the vortex flow in the \ha\ and \ca lines reflects the different formation heights and mechanisms of the two lines in the solar atmosphere. It is consistent with the behaviour of a clockwise rigidly rotating logarithmic spiral flow with a swaying motion that is  highly perturbed, however, by nearby flows associated with fibril-like structures. A radial expansion of the spiral flow has also been revealed with a mean velocity of $\sim$3\kms, and a first rough estimate of its rotational period in the range of 200 to 300~s has been provided. Concerning the long duration of the vortex flow, \cite{tzio:2018} were not able to conclusively infer from the analysed observations if it results from a) the dynamics of a central swirl acting as an ``engine'' for the whole vortex flow, b) the combined action of several individual smaller swirls that is further assisted by nearby flows, or c) a hydrodynamically driven vortex flow that has never been considered in literature so far. For the substructure, they reported that it is unclear if it exists because of the physical presence of individual intermittent and recurring swirls or if it is a manifestation of wave-related instabilities within a large vortex flow.

\section{Transverse and rotational motions}
\label{transrot}

Two main components of the vortex flow dynamics were identified in Paper~I: a rotational motion, and a superimposed swaying (transverse) motion. These two motions ideally occur on planes perpendicular to the main axis of a vortex structure that extends (nearly) vertically with height. Because the structure lies at $\sim$38\degr\ south and $\sim$10\degr\ west on the solar disc, some crosstalk between the two motions is expected as a result of projection effects. However, this crosstalk is expected to be very small because the two velocities related to these two motions are ideally almost perpendicular to each other.

\begin{figure}
\centerline{
\resizebox{\hsize}{!}{\includegraphics{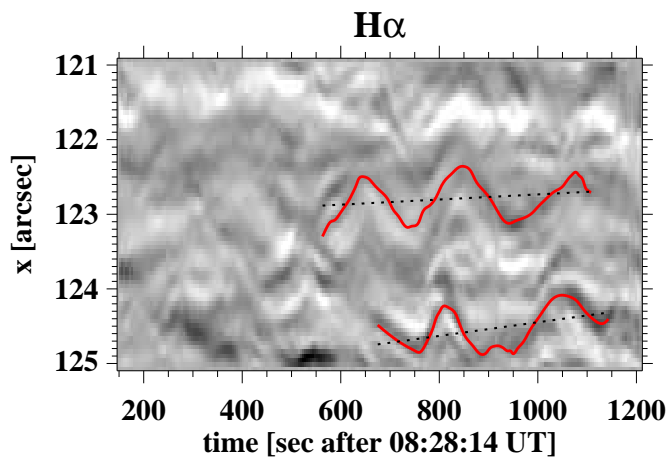}}}
\caption{Time-slice contrast-enhanced intensity along a horizontal slit passing from the approximate visual centre of the swirl found within the red circle in \fig{fig1}. This figure corresponds to a cut-out of the respective time-slice panel (top second panel) of Fig. 9 in Paper I. The contrast-enhanced intensity corresponds to the \ha\ line-centre intensity convolved with the standard Interactive Data Language (IDL) {\em emboss} function, a procedure that helps to locate edges in an image. The overplotted red lines follow the temporal evolution of two different visually identified spiral arms of this swirl, and the dotted black lines indicate the linear trend of their radial motion.} \label{fig3}
\end{figure}

The swaying (transverse) motion is clearly visible in the time-slice images of Fig.~9 in Paper I. In conjunction with the toy model of a swaying clockwise-rotating logarithmic spiral (see Fig.~11 of the same paper), this motion appears as sinusoidal evolving patterns, mostly within the vortex flow, but also in the ROI. Visual tracking of these patterns in both investigated directions suggests a period for this swaying motion of the order of 200--220~s (e.g. see the time difference between minima or maxima of sinusoidal structures in the top second panel of that figure). We tried to quantify this period by identifying two of these evolving structures that correspond to expanding spirals of the swirl found within the red circle in \fig{fig1} (\fig{fig3}, red lines). A spectral analysis of the de-trended spatial position of these structures with time gives periods of 205 and 217~s for the upper and lower curve, respectively, which are close to the visually derived periods. Because the displacement amplitude is $\sim$0.8\arcsec, this results in a mean velocity for the swaying motion of the order of $\sim$5.5\kms. Furthermore, from the linear trend of the position evolution, we estimate radial expansion velocities for these swirl spirals of 0.25 and 0.66~\kms, respectively, which are both lower than the derived overall radial expansion velocity of 3~\kms\ reported in Paper~I. This suggests an expansion within the vortex flow that is not homogeneous, and accordingly, a divergence from a rotating ideal logarithmic spiral.

\begin{figure}
\centerline{\resizebox{1.\hsize}{!}{\includegraphics{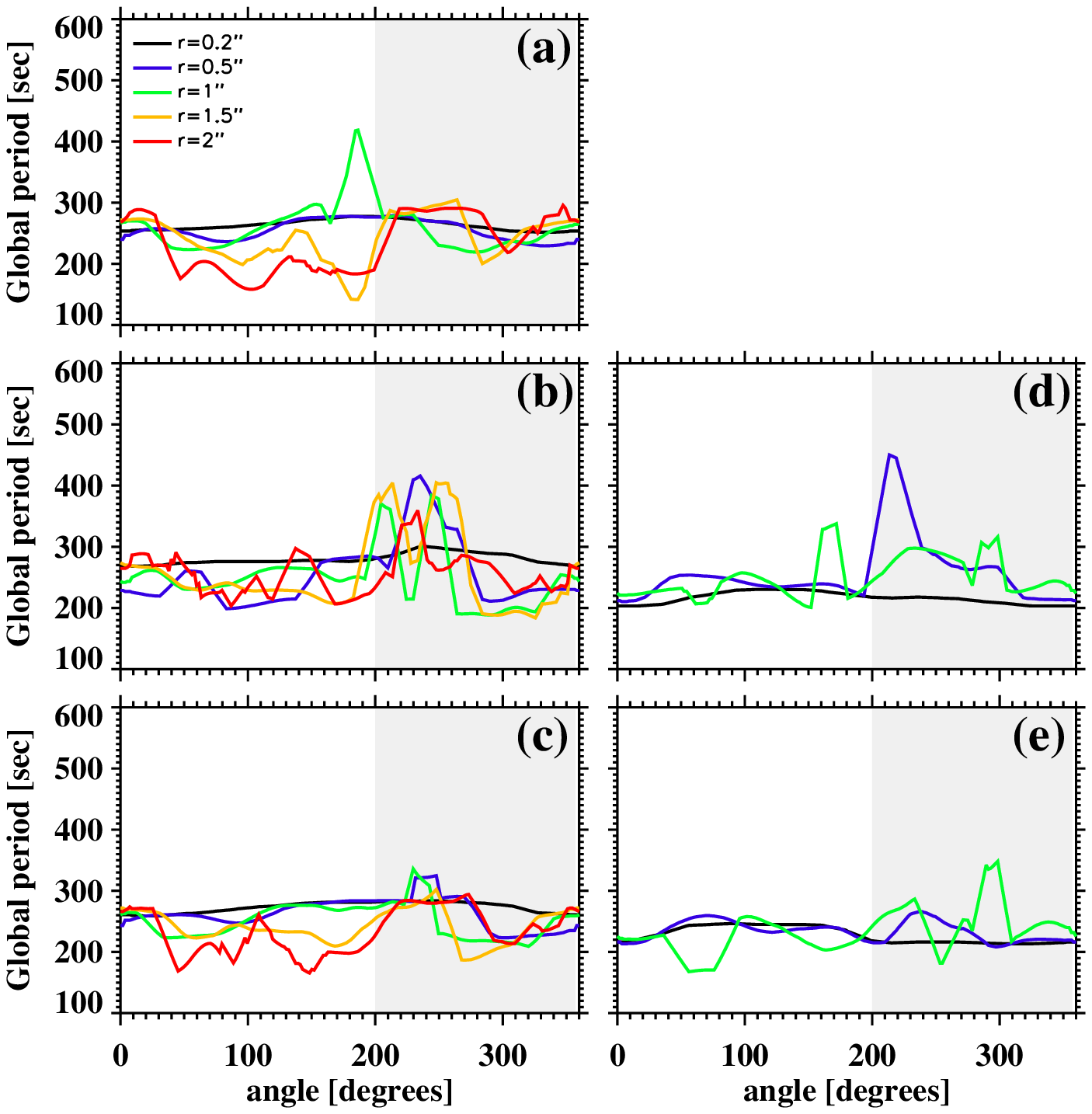}}}
\caption{Rotation periods (below 500~s) derived for the \ha\ Doppler velocity time-series (panel a) and intensity time-series at \ha\ line centre (panel b), \ha$-$0.26 \AA\ (panel c), \ca line centre (panel d), and \ion{Ca}{ii}-0.055 \AA\ (panel e) for the second observing period (08:28 UT--09:16 UT) as a function of angle and for different values of radius $r$ from the approximate centre of the vortex flow (the respective colours are indicated in the top panel). Radii extend only up to 2$\arcsec$\ for \ha, and for \ca are limited up to 1$\arcsec$\ because only the central swirl and within this radius is visible in this line (see Sect.~\ref{over}). Angles are measured in a clockwise direction starting from position 0, which is clearly marked in \fig{fig1} to the right of the vortex flow. All curves have been smoothed with a running average corresponding to $\sim$20\degr\ to remove spurious spikes.  Grey-shaded areas indicate angles where the vortex flow is mostly affected by the dynamics of nearby fibril-like flows. }\label{fig4}
\end{figure}

The rotational motion within the vortex area is clearly visible in the time-angle slices of Figs.~12 and 13 in Paper I. In conjunction with the toy model of a swaying clockwise-rotating logarithmic spiral (see Fig.~11 of the same paper), this motion appears as periodically alternating black and white stripes, whose period corresponds to the rotation of the vortex flow. In Paper~I we provided a first visual estimate of this rotation period and found it to lie within the range of 200--300~s. To quantify it, we performed a spectral analysis along horizontal cuts traced in Figs.~12 and 13 of Paper~I, which  correspond to \ha$-$0.26 \AA\ intensity, \ha\ Doppler velocity, and \ion{Ca}{ii}$-$0.055 \AA\ intensity variations, as well as the corresponding line centre \ha\ and \ca intensity variations, along different angles and for circular slits in different radii from the vortex flow centre. Derived results are presented in \fig{fig4} and Table~\ref{table1}. As discussed in Paper~I, for radii $r \ge 1\arcsec$ complete rotation is disrupted by some persisting dark features (e.g. \ha\ Doppler velocity curve in \fig{fig4}a around 180\degr\ for $r=1\arcsec$), the presence and dynamics of some fibril-like structures (mainly for angles larger than 200\degr), or some other internal or external forcing, therefore we restrict our analysis to angles $\la$200\degr.
For \ca the analysis is limited within radii $r \le 1\arcsec$  because the vortex flow is not visible further out (see Paper I). The acquired results suggest that for \ha\ Doppler velocity and intensity variations, the rotation periods are of the same order and to some extent qualitatively similar for angles that are not affected by the dynamics of different nearby flows (not grey-shaded areas in \fig{fig4}). The similarity reflects the association of upwards velocities with dark, absorbing vortex-related structures and is more clearly visible when panels a and c of \fig{fig4} are compared because the \ha$-$0.26 \AA\ intensity variations include a substantial upwards velocity contribution. For the smaller radius of $r = 0.2\arcsec$, the derived rotation period in both lines, although different, is rather constant regardless of the direction (angle) and equal to $\sim270$~s for \ha\ and $\sim215$~s for \ion{Ca}{ii}. For larger radii, the range of the rotation periods with direction increases with distance from the vortex centre, while as  noted above, clear deviations exist for angles larger than 200\degr\ (\fig{fig4}). These increasing ranges with radius also reflect deviations from a rotating ideal logarithmic spiral.

The derived rotation periods given in Table~\ref{table1} are much shorter than the 35 min period reported by \cite{Bonet:2010} for proper motions of individual vortices.
If they described a circular motion, they would correspond to horizontal tangential velocities of 3 to 45\kms\ (with a mean of 18\kms) that increase with radial distance from the vortex centre at a particular height corresponding to the wavelength of the line.

We note that the results of Table~\ref{table1} and \fig{fig4} suggest, at least for non-perturbed flows (angles $\la$200\degr) and for the common range of radii up to 1\arcsec, an increase of rotational period with height as the investigated line centre and near line-centre \ion{Ca}{ii} wavelengths form lower than the \ha\ ones. Moreover, for directions that are not disrupted by nearby flows (e.g angles $\la$200\degr), at least for \ha$-$0.26 \AA\ and more clearly for \ha\ Doppler velocity, the rotational period seems to decrease with radial distance from the vortex flow centre (despite the respective errors, which increase with radius, reported in Table~\ref{table1}). These two findings are further analysed and discussed in Sect. ~\ref{oscibeh}.

We also checked the oscillatory behaviour of the variations in the mean upwards velocity of vortex-related structures and the area they occupy (shown in Fig.~8 of Paper~I). Both quantities should be affected by the rotational motion of the vortex flow and not by transverse motions that move the vortex structure as a whole. A spectral analysis for the second observing interval (08:28 UT -- 09:16 UT) indeed yields a period of $\sim$280~s for both quantities; this is close to the derived rotation periods for \ha\ line-centre intensities and Doppler velocities.

\begin{table*}
\caption{Derived mean rotation period and respective error (standard deviation) as a function of radii from the vortex flow centre, corresponding to the spectral analysis panels of \fig{fig4} for angles ranging from 0\degr\ to 200\degr.}
\label{table1}
\centering
\begin{tabular}{l c c c c c}
\hline\hline
Radius  & \multicolumn{5}{c}{Rotation period (in s)} \\
            & \ha\ line centre  & \ha$-$0.26 \AA  & \ha\ Doppler velocity & \ca & \ion{Ca}{ii}$-$0.055 \AA \\
\hline
0.2\arcsec  & 274.6$\pm$3.2     & 271.1$\pm$8.5   & 264.4$\pm$8.1   & 210.9$\pm$8.2  & 217.7$\pm$3.2 \\
0.5\arcsec  & 237.5$\pm$28.7    & 264.6$\pm$12.6  & 257.9$\pm$14.8  & 240.4$\pm$42.0 & 233.5$\pm$14.3 \\
1\arcsec    & 251.3$\pm$13.8    & 254.5$\pm$20.0  & 271.2$\pm$46.7  & 234.1$\pm$12.3 & 220.9$\pm$25.0\\
1.5\arcsec  & 241.2$\pm$28.9    & 237.9$\pm$16.8  & 224.6$\pm$34.8  &  -             &  - \\
2\arcsec    & 246.6$\pm$28.4    & 213.4$\pm$30.5  & 203.7$\pm$37.2  &  -             &  - \\
\hline
\end{tabular}
\end{table*}

\begin{figure*}
\centerline{\resizebox{1.\hsize}{!}{\includegraphics{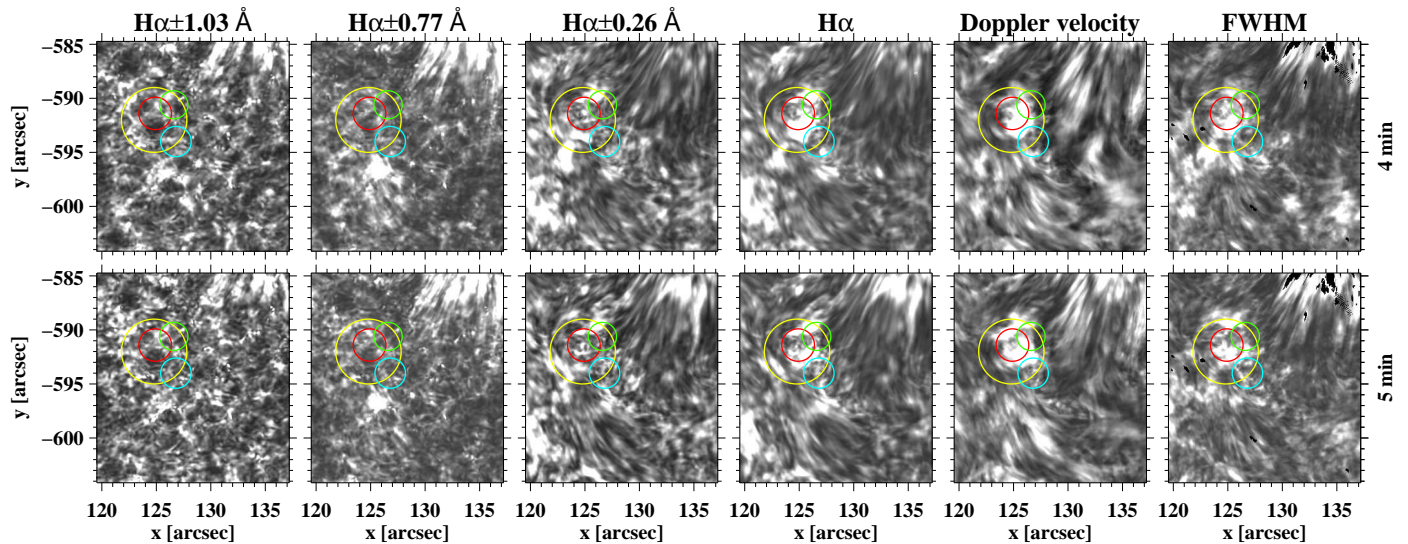}}}
\caption{2D power maps at the 4 and 5 min period bands (top and bottom row, respectively)
in the \ha$\pm$1.03 \AA, \ha$\pm$0.77 \AA, \ha$\pm$0.26 \AA, \ha\ line centre, \ha\ Doppler velocity, and FWHM for the second observing period (08:28 UT -- 09:16 UT). The two period bands of 1 min width are centred around the respective periods, while the greyscaling in each panel is within $\pm2\sigma$ from the respective mean power value of the whole image, with $\sigma$ corresponding to the respective standard deviation. The overplotted yellow, red, green, and cyan circles indicate the location of the analysed conspicuous vortex flow and of the smaller swirls indicated in \fig{fig1}.} \label{fig5}
\end{figure*}

\section{Oscillatory behaviour at different heights}
\label{res}

Below we present an analysis of the oscillatory behaviour of the intensities within the ROI and the vortex flow under study at different wavelengths along the two observed line profiles, that is, at different heights of the solar atmosphere. We furthermore investigate oscillations in \ha\ Doppler velocity and also oscillations in \ha\ FWHM because the latter have been linked with the presence of torsional Alfv\'{e}n waves \citep{Jess:2009}.
Derived global power spectra, resulting from a wavelet analysis (see Sect.~\ref{meth}) at each pixel within the ROI were used to construct two-dimensional (2D) power maps and determine the dominant oscillating period(s) and their behaviour.

We note that in order to minimise the effect of Doppler velocities on intensities taken on either wing of the \ha\ or \ca profiles at a wavelength position $\Delta\lambda$ from the line centre, for the analysis hereafter (unless stated otherwise), we used their wavelength average (e.g. \ha$\pm$0.26 \AA\ is the average intensity at wavelengths -0.26 \AA\ and +0.26 \AA).

\subsection{Power maps}
\label{pmaps}

\begin{figure}
\centerline{\resizebox{1\hsize}{!}{\includegraphics{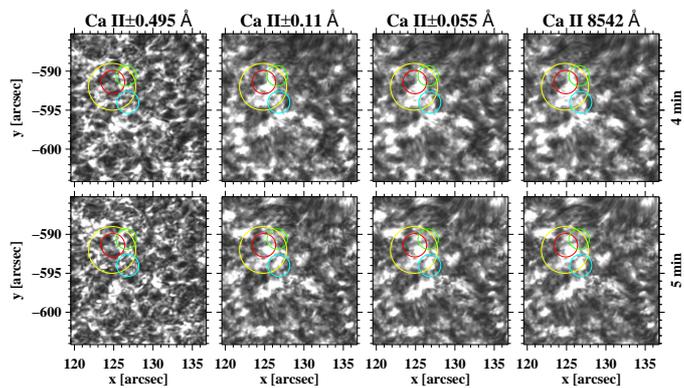}}}
\caption{Similar to \fig{fig5} 2D power maps for \ion{Ca}{ii}$\pm$0.495 \AA, \ion{Ca}{ii}$\pm$0.11 \AA, \ion{Ca}{ii}$\pm$0.055 \AA,\ and the \ca line centre. For details concerning the period bands, overplotted colour circles, and greyscaling, see the caption of \fig{fig5}.} \label{fig6}
\end{figure}

We constructed 2D power maps in two period bands for the second observing period (08:28 UT -- 09:16 UT). These power maps represent the sum of the power over all periods in the period bands centred at 4 and 5 min with a width of 1 min and are shown in Figs.~\ref{fig5} and \ref{fig6} for the \ha\ and \ca lines, respectively.
Different wavelengths in these lines represent in principle different heights of formation within the solar atmosphere, therefore these maps provide an insight into the vertical distribution of power and consequently, into oscillations at different heights. We recall that as discussed in Paper~I, the \ha\ line centre has a wide formation height range that is centred slightly above $\sim$1.5~Mm \citep[][see their Fig.~4]{Leen:2012}, with some significant contributions also from lower heights close to 0.4~Mm. In contrast, the \ion{Ca}{ii} line centre is formed at a narrower height range that is centred at $\sim$1 Mm above the photosphere \citep[][see their Fig.~4]{Leen:2009}. \ha$\pm$0.26$\,${\AA} is mainly formed somewhere between $\sim$1.2~Mm (the formation height of \ha$\pm$0.35$\,${\AA}) and the  \ha\ line-centre formation height \citep[see][their Fig.~6]{Leen:2006}, while \ion{Ca}{ii} near line-centre wavelengths are formed very close to the respective line-centre formation height within $\sim$200~km below. The wing wavelengths of the \ha\ and \ion{Ca}{ii} line profiles are formed far lower at low chromospheric and photospheric heights.

The \ha\ intensity power maps at line centre and close to line-centre wavelengths in \fig{fig5} show enhanced power in both bands within the vortex flow area, which seems to be higher in the 4 min band. Clear enhancements are also visible in the \ha\ Doppler velocity and FWHM power maps. The \ion{Ca}{ii} intensity power maps of \fig{fig6} exhibit a similar behaviour to the \ha\ ones: enhanced power in the 4 to 5 min range (higher in the 4 min band) that is visually lower close to the photosphere ($\pm$0.495 \AA) than in nearby quiet-Sun regions and becomes more enhanced farther up in the chromosphere (i.e. at near line centre and line-centre wavelengths).

The power enhancement within the vortex area is not homogeneous. All higher chromospheric diagnostics wavelengths (line centre and near line-centre \ha\ and \ion{Ca}{ii} intensities, \ha\ Doppler velocity, and FWHM) show that it is mainly enhanced within the swirl denoted with a red circle, while no obvious enhancement is visible in lower chromospheric or photospheric diagnostics wing wavelengths (\ha$\pm$0.77 \AA, \ha$\pm$1.03 \AA,\ and  \ion{Ca}{ii}$\pm$0.495 \AA). Enhancement is also visible within the swirl that is denoted with a cyan circle, especially in the \ion{Ca}{ii} higher chromospheric diagnostics. The extremely dynamic and intermittent nature of the vortex flow results in a distribution of the enhanced power within the vortex area that does not show any particularly pronounced morphology. However, some hints of vortex-like structures are faintly visible within the swirl corresponding to the red circle, while a persistent structure, already noted in Paper~I, towards the left border of the vortex flow area, appears as a dark feature because of lack of power in that location.

It is worth noting that power enhancement, especially in the 5 min band, is also seen within the fibril-like structures (top right area of the ROI) in all \ha\ wavelengths. In \ion{Ca}{ii} no particular pattern of such power is visible as related to these structures because the formation mechanism and height of this line are different from those of the \ha\ line and the height location of these structures (see Paper~I). The power enhancement in fibril-like structures is related to the magnetic field topology \citep{Konto:2010a,Konto:2010b} and the decrease in cut-off frequency (see below) caused by the presence of inclined magnetic fields \citep{Mich:1973,Suem:1990}. Hence, the similar power behaviour within the vortex area and the area of fibril-like structures could imply that the vortex area likewise is a magnetically supported structure.

We note that the respective power maps in both lines for the first observing period (07:32 UT -- 08:21 UT) indicate a similar behaviour to the description above. The power enhancement is weaker, however, and agrees with the previously reported differences in the dynamics of the vortex flow during the two observing intervals.

\begin{figure}
\centerline{\resizebox{0.7\hsize}{!}{\includegraphics{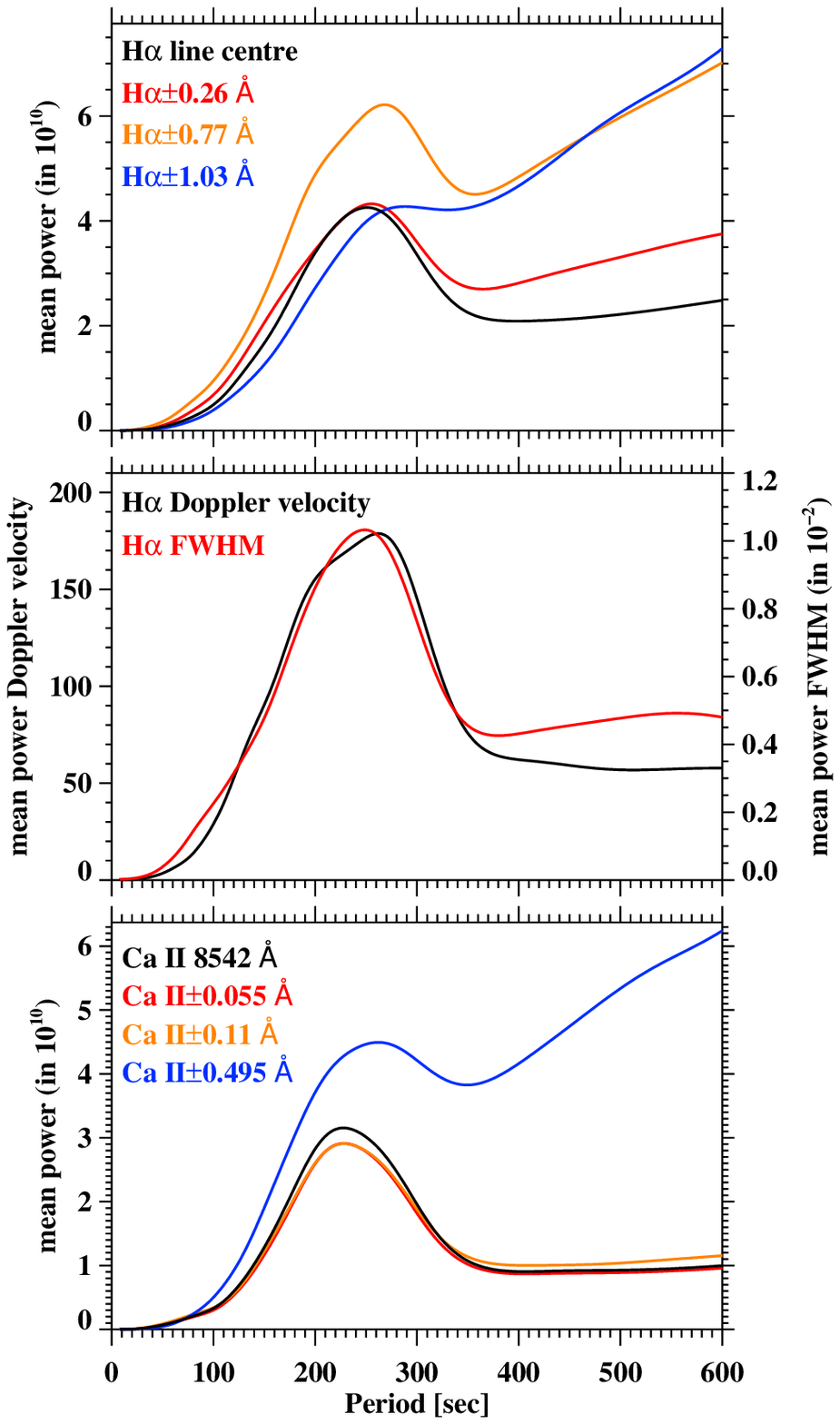}}}
\caption{Mean oscillatory power per period within the vortex area for wavelengths along the \ha\ and \ca line profiles (top and lower panels) and \ha\ Doppler velocity and FWHM (middle panel). The respective mean power errors over all periods are 0.94, 1.14, 1.77, and 1.6 (in units of 10$^{10}$) for \ha\ line centre to wing wavelengths, 34.2 and 0.18$\times10^{-2}$ for \ha\ Doppler velocity and FWHM, respectively, and 0.51, 0.47, 0.49, and 1.57 (in units of 10$^{10}$) for \ion{Ca}{ii} line centre to wing wavelengths.
} \label{fig7}
\end{figure}

Figure~\ref{fig7} shows the mean power per period within the vortex area for different wavelengths along the \ha\ and \ca line profiles (upper and lower panels), as well as the mean powers per period of the \ha\ Doppler velocity and FWHM (middle panel). In the 200--300~s range, all power curves have a peak at $\sim$220-250s. It is interesting to note that although the vortex flow is not visually discernible in the \ha\ wing wavelengths or in the \ca line, there is considerable power in the vortex area in this period range in all atmospheric heights. For periods higher than 340~s, the power is higher in the wing wavelengths than in the line centre or near line centre ones and increases linearly as a function of period for all wing wavelengths in both lines. The power in \ha$\pm$0.77 \AA\  and \ion{Ca}{ii}$\pm$0.495 \AA, mapping the upper photosphere, is higher than the power in all other wavelengths mapping the lower photosphere or the chromosphere for all periods. We note that \ion{Ca}{ii} near centre and line-centre wavelengths show an identical power distribution that  mainly reflects the quite small height difference at which they form. The \ha$\pm$1.03 \AA\ power clearly deviates from the apparent decrease in power from lower to higher atmospheric layers. However, we note that there only exist very faint but highly intermittent signatures of the vortex flow in \ha$\pm$1.03 \AA\ (see Paper I), while the errors associated with this power and the power of \ha$\pm$0.77 \AA\ (see caption of \fig{fig7}) are very large.
The derived power decrease with height in both lines is consistent with the expected behaviour for upwards-travelling evanescent waves.

\begin{figure}
\centerline{\resizebox{1\hsize}{!}{\includegraphics{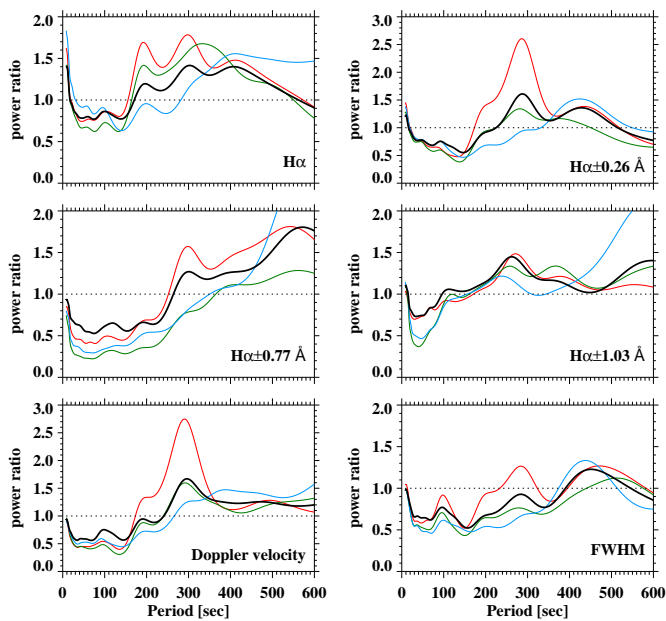}}}
\caption{Ratio between average power within vortex-related areas and average power within the selected QSR (see \fig{fig1}) for different \ha\ wavelengths, \ha\ Doppler velocity, and \ha\ FWHM. The black line corresponds to the ratio for the whole vortex area, and the colour lines correspond to ratios within swirls that are denoted with the same colours in \fig{fig1}.
} \label{fig8}
\end{figure}

\begin{figure}
\centerline{\resizebox{1\hsize}{!}{\includegraphics{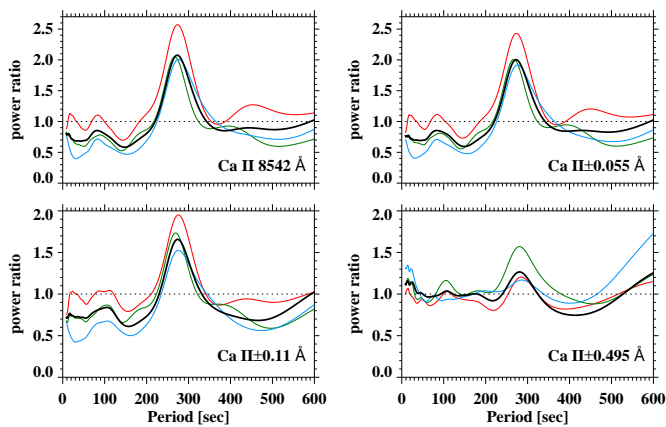}}}
\caption{Ratios for \ion{Ca}{ii} wavelengths, similar to \fig{fig8}. For details concerning different colour lines,
see the caption of \fig{fig8}.} \label{fig9}
\end{figure}

To better quantify the differences in the power behaviour at different periods within the vortex area compared to nearby regions in the ROI, we examined ratios
between average power within vortex-related areas and average power within the selected QSR. The obtained ratios are shown in \fig{fig8} for \ha\ and in \fig{fig9} for \ion{Ca}{ii}.
The two figures show that for both lines, the power is generally more enhanced in the QSR than in the vortex area for periods shorter than $\sim$200~s (ratio below one). For periods >200~s and within the range 200$-$350~s, the power is more enhanced within the vortex area than in the QSR (ratio above one) not only at photospheric, but also at chromospheric heights. We note that the picture is clearer for \ion{Ca}{ii} ratios at different wavelengths than the \ha\ ones, reflecting the more intricate formation physics of the \ha\ profile, which receives contributions from different heights at certain wavelengths \citep[e.g. \ha$\pm$0.26 \AA, see Fig. 6 of][]{Leen:2006}.
We further discuss Figs.~\ref{fig7}, \ref{fig8}, and \ref{fig9} and how the power behaviour relates to the presence of waves in Sect.~\ref{varperwav}.

\begin{figure*}
\centerline{\resizebox{0.9\hsize}{!}{\includegraphics{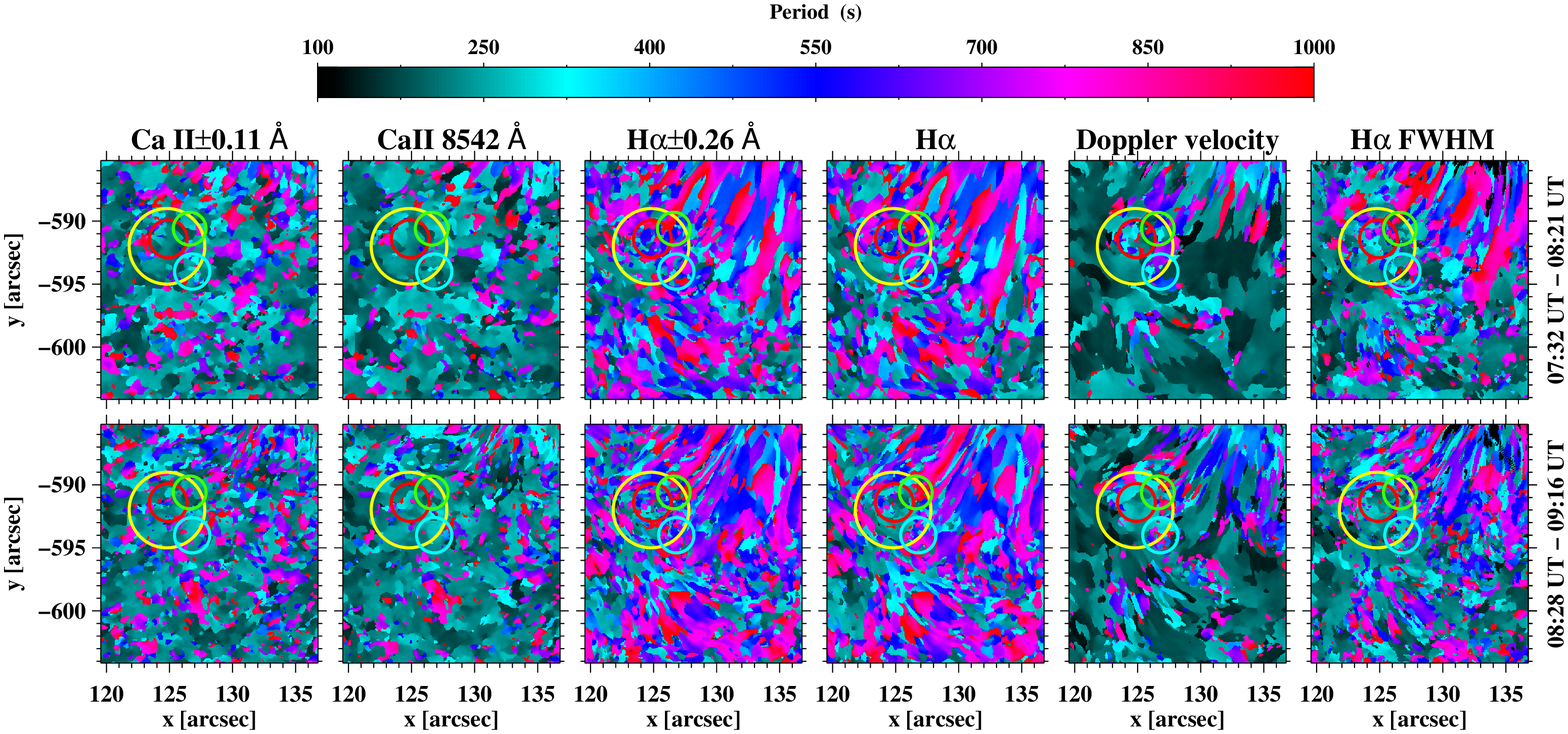}}}
\caption{Maps of the dominant period derived from the global power spectrum within the ROI during both observing intervals (upper and lower rows, respectively) for \ion{Ca}{ii}$\pm$0.11 \AA, \ca line centre, \ha$\pm$0.26$\,${\AA,}\ and \ha\ line centre intensity variations, and for the \ha\ Doppler velocity and \ha\ FWHM variations.
The overplotted yellow, red, green, and cyan circles indicate the location of the analysed conspicuous vortex flow and of the smaller swirls indicated in \fig{fig1}.}\label{fig10}
\end{figure*}

The power within the vortex area is mostly dominated by the swirl denoted by a red circle in \fig{fig1}. This is clearly visible in the 2D power maps (Figs.~\ref{fig5} and \ref{fig6}) and also in the identical behaviour of the respective ratios of the average power (Figs.~\ref{fig8} and \ref{fig9}).
We recall that the vortex area comprises the swirl within the red circle and only partly the swirls within the cyan and green circles marked in \fig{fig1}. The latter two swirls do follow the general trend, with some differences mostly in the complex \ha$\pm$0.26 \AA\ and \ha$\pm$0.77 \AA\ wavelengths that also stem from their behaviour, which is more intermittent and erratic than the behaviour of the swirl within the red circle (see Paper~I). These differences and similarities between the behaviours of the three swirls and the vortex flow as a whole seem to imply that the swirl within the red circle might be the core of the vortex flow, and the other two might mostly be caused by the dynamics of the vortex flow.

\begin{figure*}
\centerline{\resizebox{1.\hsize}{!}{\includegraphics{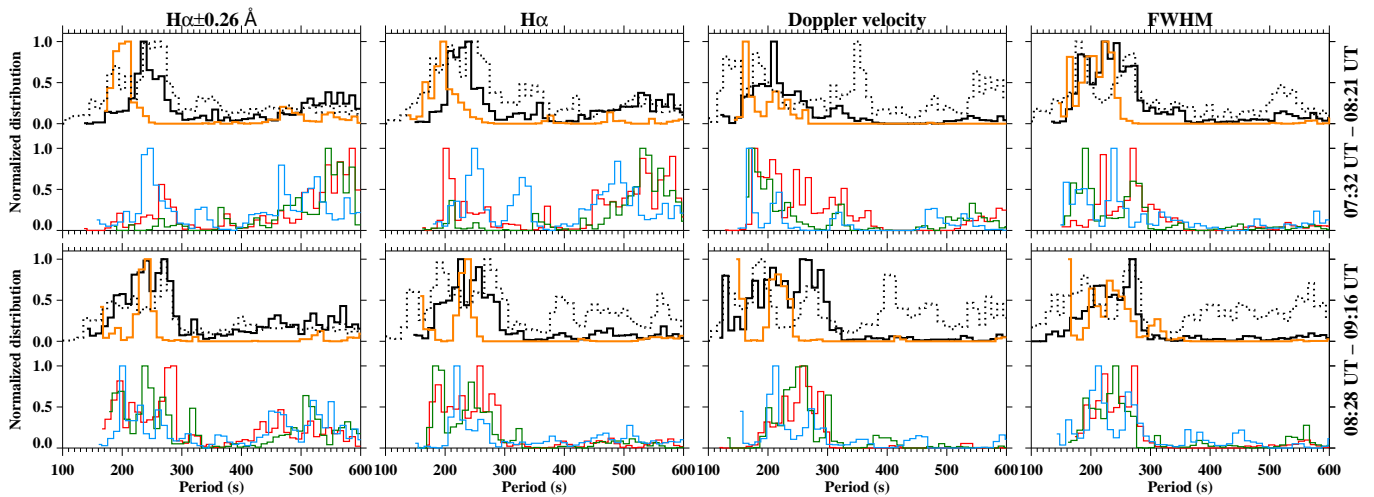}}}
\caption{Histograms of the dominant periods (solid lines) of the global power spectrum below 1000~s (normalised to their maximum) within the vortex area (black line, area within the yellow circle in \fig{fig10}), the swirls within the red, green, and cyan circles in \fig{fig10} (red, green, and cyan lines, respectively) and the QSR (orange line, white rectangle in \fig{fig1}) for the maps in \fig{fig10}. The dotted lines denote the distributions of the second dominant period of the global power spectrum below 1000~s within the vortex area. We recall that the vortex area distribution contains the entire distribution corresponding to the swirl denoted by a red circle, and parts of the distributions corresponding to swirls denoted by the cyan and green circles in \fig{fig10}.}\label{fig11}
\end{figure*}

\begin{figure}
\centerline{\resizebox{1.\hsize}{!}{\includegraphics{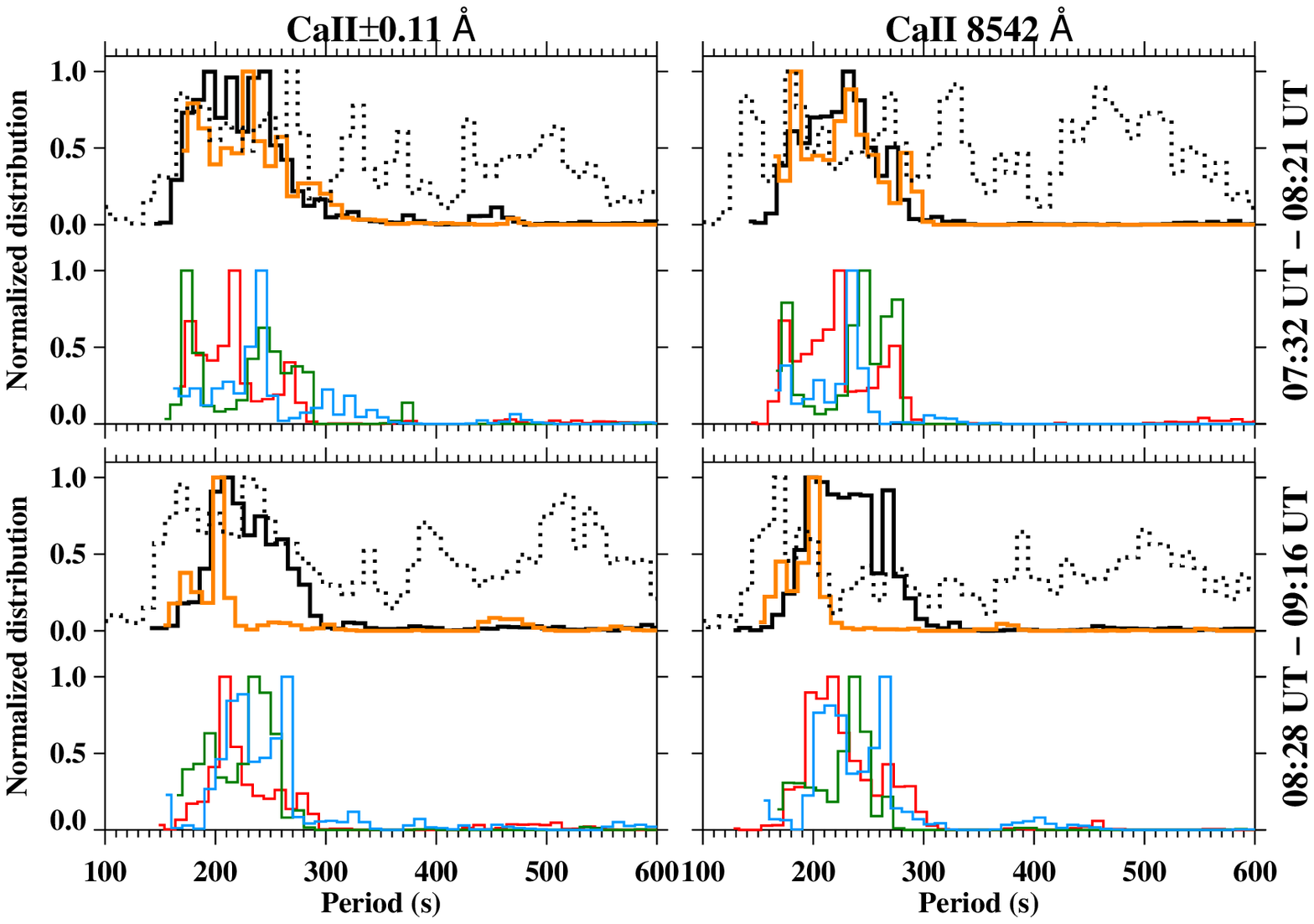}}}
\caption{Normalised histograms of the dominant periods, similar to \fig{fig11}, for the maps of \ion{Ca}{ii}$\pm$0.11 \AA\ and \ca line centre of \fig{fig10}. For the description of different colours and lines see the caption of \fig{fig11}.}\label{fig12}
\end{figure}

\subsection{Dominant oscillation periods}
\label{osci}

From the derived global power spectra at each pixel of the ROI we found the dominant period below 1000 s and constructed respective 2D maps (\fig{fig10}) and the corresponding histogram distributions (Figs.~\ref{fig11} and \ref{fig12}) for intensity variations at certain wavelengths along the \ha\ and \ca profiles, as well as for \ha\ Doppler velocity and \ha\ FWHM temporal variations. The threshold of 1000~s was chosen to avoid high power and consequently, periods stemming from edge-effects due to our finite time-series. We recall that the dominant period is the peak of a usually wider period distribution (e.g. see \fig{fig2}), therefore it is an indicative period, while derived global power spectra in \ca are generally broader around the dominant period than those of \ha.

The \ha\ dominant-period maps of \fig{fig10} indicate that the vortex area generally oscillates differently than the rest of the ROI. Oscillations at locations outside the borders of the vortex show either lower periods (e.g. QSR of \fig{fig1} and area around) that coincide with the expected 3 min chromospheric oscillations, or higher periods (e.g. areas with fibril-like structures at the top and top right of the vortex area) that coincide with the dynamics and lifetimes of such structures \citep{Tsir:2012,Konto:2014}. However, within the vortex area, the oscillatory pattern is not homogeneous. Although the larger part of the vortex area does not seem to differ substantially from the quiet-Sun regions as it oscillates at only slightly higher periods, the locations of the three individual swirls, and mostly that of the swirl within the red circle, exhibit somewhat higher periods than the rest of the vortex area (see also the respective histograms of \fig{fig11} and related discussion).

As first indicated by \fig{fig2} and now further suggested by \fig{fig10}, both intensity (especially at \ha$\pm$0.26 \AA) and Doppler velocity dominant-period maps and to some degree the FWHM maps exhibit a similar oscillatory behaviour, at least within the individual swirls and to a large extent of the rest of the vortex area, especially its south-eastern part. This similarity between intensity and Doppler-velocity oscillations probably results from the correlation between dark absorbing features that dominate the intensity within the vortex and upwards velocities (see Paper~I).

\ca line centre and near line centre dominant-period maps are almost identical, as was suggested by the respective power behaviour in Figs.~\ref{fig7} and \ref{fig9} before. The maps of \fig{fig10} show that some differences exist in the observed oscillatory pattern when compared to those of \ha\ (see also the respective histogram distributions of \fig{fig12} and related discussion). The oscillations both in the vortex area and the locations of individual swirls seem to be more homogeneous and coherent in this line than in the respective \ha\ oscillations, while high dominant oscillation periods seem to be absent. This probably reflects a) the less violent dynamics in the \ca line (due to its lower formation heights) where the swirl substructure is hardly visible, with the exception of the central swirl within the red circle, b) the funnel-like expansion of the vortex flow itself with height, and c) the absence of interference with the dynamics of nearby to the vortex flow area fibril-like structures that seem to lie mostly higher than the formation heights of the \ca line, as indicated in Paper~I.

We note that some differences exist between the oscillatory behaviour during the two observing intervals for the \ha\ line, as mostly \fig{fig11} clearly indicates, which are not seen or are less pronounced in the \ca line, with the exception again of the area of the swirl within the red circle. These differences arise from the characteristics of the vortex flow itself, which is mostly visible after the second half of the first observing period (see Paper~I) .

We quantified the oscillatory behaviour shown in the dominant-period maps above with the corresponding histograms of the period distribution shown in \fig{fig11} for \ha\ and \fig{fig12} for \fca.
For both lines, the period distributions within the vortex flow area reveal dominant periods in the range of 3 to 5 min that mostly peak around 4 min. This agrees with the behaviour observed and discussed in the power maps and figures of Sect.~\ref{pmaps}. Furthermore, most of the distributions, especially during the second observing interval of 08:28 UT--09:16 UT, seem to have two peaks around 220 and 270 s (the latter period is also clearly visible at least in the global power spectra of intensity and velocity in \fig{fig2}). These peaks are also evident in the distributions of the second dominant period, especially for \ha, further indicating their significance and suggesting that the two dominant periods alternate most of the times between these two values. Of the three identified individual swirls, it is mainly the swirl within the red circle that follows the distribution of periods of the whole vortex area, in particular during the less erratic second observing interval. This is more evident in the \ca line distributions, where it is the only clearly observed chromospheric swirl, further supporting the argument presented in Paper~I and above, that this swirl probably serves as a ``central engine'' for the entire vortex flow. Although the other swirls clearly show periods within the 3 to 5 min range, they exhibit a more erratic period distribution that reflects their highly intermittent and far more complex dynamics.

When we compare the \ha\ and \ion{Ca}{ii} distributions with the respective distributions resulting from the spectral analysis of the QSR of \fig{fig1} (orange line in Figs.~\ref{fig11} and \ref{fig12}), we note, as we discussed in Sect.~\ref{pmaps} and related figures, that the QSR oscillates differently than the vortex area and most of the individual swirls. It exhibits shorter periods that most of the times are closer to the expected 3min chromospheric oscillations and lie within the range of 180--220~s.

\section{Discussion of the oscillatory behaviour}
\label{oscibeh}

The derived oscillations in such a complex and dynamic vortex flow reflect the cumulative action of different components, such as rotation, radial expansion, swaying (transverse) motions, perturbations from nearby flows, and propagating waves. The rotation of the structure, however, may have the greater influence. In the following subsections we discuss the main characteristics of the derived oscillatory behaviour and try to gain insight into the dynamics of the vortex flow.

\subsection{Variations in oscillatory power within the vortex flow, comparison to the QSR power, and relationship to waves}
\label{varperwav}

A wavelet spectral analysis permeated the construction of 2D power maps in several wavelengths along the \ha\ and \ca profiles, as well as of \ha\ Doppler velocity and FWHM. The 2D power maps show no particular substructure because the vortex flow is extremely dynamic, but they indicate enhanced power and a faint vortex-like signature around the location of the swirl within the red circle marked in \fig{fig1}. The analysis of the power behaviour with period, on the other hand, revealed that considerable oscillating power is found within the vortex flow in the range of 3 to 5 min that peaks around 4 min with at least qualitatively similar characteristics in intensities of the two investigated lines and in \ha\ FWHM and Doppler velocity. Enhanced power is, however, also visible at higher periods up to 10 min, mostly at lower, but to some extent also at higher atmospheric heights (i.e. from wing to line-centre wavelengths). The power seems to decrease in both lines from upper photospheric to chromospheric layers for periods mainly in the range of 3 to 5 min but even up to 10 min, suggesting the presence of evanescent waves.

The analysis of the power behaviour with period within the vortex flow compared to the power in the QSR reported in Sect.~\ref{pmaps} together with the dominant period analysis reported in Sect.~\ref{osci} provides evidence for a different oscillatory behaviour in these two regions. In quiet-Sun regions, the power at photospheric heights (e.g. \ha$\pm1.03\,${\AA} and further into the \ha\ wings) is dominated by the acoustic global $p$-mode waves with a period distribution that peaks at $\sim$5~min. P-modes are normally evanescent and can only propagate upwards if they have frequencies higher than the acoustic cut-off frequency $f_{\rm 0}=5.2$~mHz. The acoustic cut-off frequency generally depends on the local plasma $\beta$ and the magnetic field inclination angle $\theta$ with respect to the direction of gravity that follows the solar normal. In the presence of strong (low-$\beta$ regime) inclined magnetic fields it is modified as $\sim f_{\rm 0} \cos\theta$, permitting the otherwise evanescent $p$-modes to propagate upwards and reach higher atmospheric layers. In the case of low magnetic fields, the lowering of the acoustic cut-off frequency may be due to a different mechanism such as radiative damping in an isothermal atmosphere \citep[e.g.][]{Worr:2002}.

The QSR generally oscillates and shows more power at shorter periods closer to the 3~min chromospheric period. For periods >200~s, the power within the vortex area is greater than in the QSR  (see Figs.~\ref{fig8} and \ref{fig9}), and particularly in the range of 200--350~s. This enhanced power could also partly be attributed to either the presence of inclined fields or radiative damping that permit the upwards channeling of acoustic power. However, as power enhancement is clearly visible within the vortex area compared to the QSR for periods up to 10 min (see also the period distributions of \fig{fig11}), not only at low but also at high atmospheric heights (see \fig{fig8} and \fig{fig9}), it may suggest that this enhanced power is mainly due to other types of oscillations, including different types of waves such as magnetoacoustic (e.g. kink) or Alfv\'{e}n waves. Many works have indeed demonstrated that processes related to vortex formation, like turbulent convection or twisting motions exercised in the magnetic field concentrations, can excite different types of waves in the solar atmosphere. It has been reported that power at long periods is probably not related to acoustic oscillations, but might for instance be attributed to Alfv\'{e}nic disturbances \citep{McAteer:2002,McAteer:2003,Jess:2009} or kink waves that have a much lower cut-off frequency than the acoustic one \citep{Spruit:1981,Stanga:2015}. Oscillations and enhanced power of \ha\ FWHM have been associated with torsional Alfv\'{e}n waves \citep{Jess:2009}. Kink waves, on the other hand, lead to periodic transverse displacement of flux tubes and thus to periodic Doppler velocities in waveguides. Because these  waves also have substantial azimuthal velocities they can contribute to periodic variations of the non-thermal line widths \citep{Goos:2014}. We also recall that in Paper~I we suggested that waves might be related to the substructure within the vortex flow. This substructure appearing as individual, intermittent swirling patches could well be a manifestation of instabilities driven by the presence of torsional Alfv{\'e}n waves, as indicated by simulations \citep{Fedun:2011b}. Further analysis of the obtained results will help to gain more insight into the wave propagation within vortexes.

The presence and signatures of waves, their identification, propagation properties, and their influence in the vortex flow dynamics will be meticulously investigated in a future work where a detailed phase difference analysis will be used.

\subsection{Characteristic oscillation periods and their relation to the vortex dynamics}
\label{charper}

It is clear from the analysis presented here and in Paper~I  that among the different oscillatory components reflected in the dominant period maps, two are the most predominant related to the vortex dynamics: rotation, and a superimposed swaying motion. Although both swaying and rotation are horizontal motions that occur on planes perpendicular to the main axis of the structure, they do have significant components and therefore affect oscillations along the line of sight (LOS) because the vortex flow is situated at location S38E10. Transverse motions analysed in evolving spiral structures within the vortex flow reveal a characteristic period between 200--220~s and a mean swaying velocity of $\sim$5.5\kms, as well as variable expanding velocities lower than 1\kms, suggesting a non-homogeneous expansion diverging from a rotating ideal logarithmic spiral. The rotation of the vortex flow analysed along circular slits of increasing radii from its centre reveal rotation periods close to $\sim$270~s for \ha. For the \ca line, the respective periods are slightly shorter (see Sects.~\ref{transrot} and \ref{varperhei}). The derived rotation periods are shorter than previous estimates reported from observations \citep[35~min from][]{Bonet:2010}. Rotation also seems to modulate the mean upwards \ha\ velocity variations of vortex-related structures and the area they occupy because both yield a similar period of $\sim$280~s. The two acquired periods that correspond to swaying and rotational motions appear as significant peaks in the acquired dominant-period maps and respective period distributions within the vortex area, showing the significance of the two motions in the vortex dynamics.

\subsection{Variation in oscillation periods with height}
\label{varperhei}
Oscillations in the different wavelengths of the \ca and \ha\ line profiles represent oscillations at different heights of the atmosphere, with the \ca line centre and near line-centre wavelengths being formed lower than the corresponding \ha\ wavelengths. The period distributions in \ion{Ca}{ii} shown in \fig{fig12} seem to shift towards shorter periods when compared to the respective \ha\ period distributions shown in \fig{fig11}. This shift is even clearer when comparing in \fig{fig4} and Table~\ref{table1} the respective average periods in the two lines that were derived along circular slits of similar radii. We may conclude that these findings suggest an increase in oscillation period with height. This result is in agreement with simulations by \citet[][see their Fig. 4b]{Moll:2011b}, although for very low (close to the solar surface) atmospheric heights that indicate a move of the peak of the swirling period $T_{\rm sw}$ towards higher values with increasing height.

If periods were only associated with acoustic waves travelling in a magnetically supported vortex structure, their increase with height could be attributed to a lowering of the cut-off frequency, for example, due to the presence of more inclined fields with height, as in the simulations of swirling magnetic fields within a funnel-like expanding vortex flow  \citep[][see Figs. 1 and 2]{Wede:2014}. However, as the results of Sect.~\ref{pmaps} and discussion in Sect.~\ref{varperwav} indicate, other types of waves and oscillations may be related to the derived periods. The period increase with height might be explained in terms of preservation of angular momentum and mass. Assuming that the upwards mass ($m$) flow is constant and the vortex structure expands with height (funnel-like structure), as the findings of Paper~I suggest, this would dictate that its radius $R$ increases as we move upwards. Preservation of angular momentum ($\sim m R^2/T_{\rm rot}$) with height
would then lead to an increase in swirling period $T_{\rm rot}$, as observed. However, this is a crude assumption as a) the increase in period is clearly not of the expected order, and b) it would contradict a rigidly or quasi-rigidly rotating structure as suggested in Paper~I and in the literature \citep{Wede:2012}. Preservation of angular momentum, as would be expected in a purely hydrodynamic case, is not the main driver behind a vortex flow of this type because the oscillation characteristics we find (see previous Sect.~\ref{charper}) require the presence of a guiding magnetic field. This is mostly a magneto-hydrodynamic case where the rotation of the supporting magnetic field and its subsequent dynamics together with the action of the Lorentz force in the moving plasma \citep[e.g.][but for giant tornados]{Luna:2015} provide the necessary set-up that defines the trajectories along which the vortex-associated plasma moves.

\subsection{Variation in oscillation periods with radius at a certain height}
\label{varperrot}
Oscillations in the different wavelengths of \ca and \ha\ line profiles along circular slits, like those presented in \fig{fig4} and Table~\ref{table1}, represent mostly rotational motions, as discussed in Sect.~\ref{transrot}. As we noted in Sect.~\ref{transrot}, the results suggest that for directions where the vortex flow is not disrupted by the presence of other structures, such as the fibril-like structures, the oscillation period generally seems to decrease, at least for \ha, with increasing distance from the vortex centre. Although the errors of derived periods increase significantly with distance (see Table~\ref{table1}), this behaviour would imply a deviation of the vortex structure from a rigid rotation to a quasi-rigid rotation as we move outwards.

Simulations \citep[][see their Fig. 3f]{Wede:2012} suggest that for a certain height at the atmosphere, the horizontal velocity, which is associated with flows along spiral trajectories, increases with distance from the centre of the vortex flow up to a cut-off radius that represents the radial extent of the vortex flow; this cut-off radius increases with height and produces a funnel-like expanding structure. If this simulated horizontal velocity were to correspond to circular motions of homogeneously distributed matter ($\varv \sim 2 \pi R / T_{\rm rot}$), its clear deviation from a linear increase with $R$ would imply an increase in period $T_{\rm rot}$ with distance from the vortex centre. This is opposite to the behaviour observed, as discussed above. It is unclear if interaction with nearby tangentially parallel flows at the right (west) borders of this clockwise vortex flow (see online movie of Paper~I) could further assist its rotation at larger radii, leading to a decrease in period as observed.

\section{Conclusions}
\label{conc}

In Paper~I we analysed the characteristics and dynamics of a 1.7 h persistent vortex flow that resembles a small-scale quiet-Sun tornado. The analysis revealed a rigidly or quasi-rigidly rotating funnel-like expanding structure with a rich substructure that manifested itself in the form of a number of recurring, intermittent smaller chromospheric swirls. The vortex flow appears to also display signatures at atmospheric heights up to the low corona. Signatures of swaying (transverse motions) and rotational motions with periods in the range of 200--300~s were found but not examined in detail. In this study we took our analysis a step further by investigating both the transverse and rotational oscillatory behaviour at one particular height, the oscillatory behaviour  at different radii from the centre of the vortex flow, and at different heights in the solar atmosphere within this vortex flow. Our findings indicate significant oscillatory power within the vortex flow related to motions and waves mainly in the range of 3 to 5~min that peaks at 4~min but extends even up to 10~min at all heights. We also found periods of 200--220~s for swaying motions and rotation periods that seem to increase with height ($\sim$270~s for \ha\ and $\sim$215 for \fca) and decrease with radius from the vortex centre. Furthermore, they indicate the presence of different types of evanescent waves, in addition to acoustic, such as magnetoacoustic (e.g. kink) or Alfv\'{e}n waves. An analysis like this has to our knowledge not been done so far, at least not in this detail.

The results we presented show that the investigation of the oscillatory behaviour and of the power spectra in different wavelengths of spectral lines is a powerful tool for providing important information regarding the nature of vortex flows.
Unfortunately, our study cannot provide any explanation for the long duration (i.e. at 1.7~h) of the vortex flow. However, it furnishes further evidence to some pertinent open questions, following the analysis of Paper~I, that we address below.

Is the vortex structure magnetically supported?
Our results still do not provide any direct evidence that could imply that the vortex structure is magnetically supported. However, as the vortex area and the area of fibril-like structures show similar power enhancements within the 5~min band that in fibril-like structures has been clearly related to the magnetic field topology \citep{Konto:2010a,Konto:2010b}, this could indirectly imply that the vortex flow is also a magnetically supported structure. Furthermore, the dominant oscillation periods and the power behaviour imply the presence of magnetoacoustic and/or Alfv\'{e}n waves. These in turn require the presence of a magnetic structure.

Is there a central swirl that powers the vortex flow? The 2D power maps indicate, especially at higher chromospheric diagnostics wavelengths (e.g. \ha\ and \ca line centre and near line centre), clear enhancements within the swirl denoted with a red circle in \fig{fig1}. Comparison of the power within the vortex area and individual swirls (Figs.~\ref{fig5} and \ref{fig6}) clearly shows a dominance of the same swirl over the vortex area. Furthermore, during the second observing interval, mainly the period distribution of the swirl within the red circle resembles and dominates the distribution of periods of the whole vortex area (Figs.~\ref{fig11} and \ref{fig12}). These findings suggest that this swirl probably serves as a ``central engine'' for the entire vortex flow.

The different oscillations revealed in this study point to the existence of various modes of MHD waves within the vortex flow. Their  generation and propagation are of particular interest because they are considered to provide a potential mechanism for the channelling of energy from the lower to the upper layers of the solar atmosphere. Further studies based on the co-spatial and simultaneous CRISP observations including a phase-difference analysis will be presented in a future work.

\begin{acknowledgements}
The authors would like to thank the International Space Science Institute (ISSI) in Bern, Switzerland, for the hospitality provided to the members of the team
on ``The Nature and Physics of Vortex Flows in Solar Plasmas''.
We acknowledge support of this work by the project ``PROTEAS II'' (MIS 5002515), which is implemented under the Action ``Reinforcement of the Research and Innovation Infrastructure'', funded by the Operational Programme ``Competitiveness, Entrepreneurship and Innovation'' (NSRF 2014-2020) and co-financed by Greece and the European Union (European Regional Development Fund). The Swedish 1-m Solar Telescope is operated on the island of La Palma by the
Institute for Solar Physics of Stockholm University in the Spanish Observatorio
del Roque de los Muchachos of the Instituto de Astrofisica de Canarias.
\end{acknowledgements}

\bibliographystyle{aa}
\bibliography{vortex2} 

\newcommand{\noop}[1]{}
\begin{thebibliography}{45}
\expandafter\ifx\csname natexlab\endcsname\relax\def\natexlab#1{#1}\fi

\bibitem[{{Attie} {et~al.}(2009){Attie}, {Innes}, \& {Potts}}]{Attie:2009}
{Attie}, R., {Innes}, D.~E., \& {Potts}, H.~E. 2009, \aap, 493, L13

\bibitem[{{Barthol} {et~al.}(2011){Barthol}, {Gandorfer}, {Solanki},
  {Sch{\"u}ssler}, {Chares}, {Curdt}, {Deutsch}, {Feller}, {Germerott},
  {Grauf}, {Heerlein}, {Hirzberger}, {Kolleck}, {Meller}, {M{\"u}ller},
  {Riethm{\"u}ller}, {Tomasch}, {Kn{\"o}lker}, {Lites}, {Card}, {Elmore},
  {Fox}, {Lecinski}, {Nelson}, {Summers}, {Watt}, {Mart{\'{\i}}nez Pillet},
  {Bonet}, {Schmidt}, {Berkefeld}, {Title}, {Domingo}, {Gasent Blesa}, {Del
  Toro Iniesta}, {L{\'o}pez Jim{\'e}nez}, {{\'A}lvarez-Herrero},
  {Sabau-Graziati}, {Widani}, {Haberler}, {H{\"a}rtel}, {Kampf}, {Levin},
  {P{\'e}rez Grande}, {Sanz-Andr{\'e}s}, \& {Schmidt}}]{Barth:2011}
{Barthol}, P., {Gandorfer}, A., {Solanki}, S.~K., {et~al.} 2011, \solphys, 268,
  1

\bibitem[{{Bonet} {et~al.}(2008){Bonet}, {M{\'a}rquez}, {S{\'a}nchez Almeida},
  {Cabello}, \& {Domingo}}]{Bonet:2008}
{Bonet}, J.~A., {M{\'a}rquez}, I., {S{\'a}nchez Almeida}, J., {Cabello}, I., \&
  {Domingo}, V. 2008, \apjl, 687, L131

\bibitem[{{Bonet} {et~al.}(2010){Bonet}, {M{\'a}rquez}, {S{\'a}nchez Almeida},
  {Palacios}, {Mart{\'{\i}}nez Pillet}, {Solanki}, {del Toro Iniesta},
  {Domingo}, {Berkefeld}, {Schmidt}, {Gandorfer}, {Barthol}, \&
  {Kn{\"o}lker}}]{Bonet:2010}
{Bonet}, J.~A., {M{\'a}rquez}, I., {S{\'a}nchez Almeida}, J., {et~al.} 2010,
  \apjl, 723, L139

\bibitem[{{Brandt} {et~al.}(1988){Brandt}, {Scharmer}, {Ferguson}, {Shine}, \&
  {Tarbell}}]{Brandt:1988}
{Brandt}, P.~N., {Scharmer}, G.~B., {Ferguson}, S., {Shine}, R.~A., \&
  {Tarbell}, T.~D. 1988, \nat, 335, 238

\bibitem[{{Cauzzi} {et~al.}(2009){Cauzzi}, {Reardon}, {Rutten}, {Tritschler},
  \& {Uitenbroek}}]{Cauzzi:2009}
{Cauzzi}, G., {Reardon}, K., {Rutten}, R.~J., {Tritschler}, A., \&
  {Uitenbroek}, H. 2009, \aap, 503, 577

\bibitem[{{De Pontieu} {et~al.}(2014){De Pontieu}, {Title}, {Lemen}, {Kushner},
  {Akin}, {Allard}, {Berger}, {Boerner}, {Cheung}, {Chou}, {Drake}, {Duncan},
  {Freeland}, {Heyman}, {Hoffman}, {Hurlburt}, {Lindgren}, {Mathur}, {Rehse},
  {Sabolish}, {Seguin}, {Schrijver}, {Tarbell}, {W{\"u}lser}, {Wolfson},
  {Yanari}, {Mudge}, {Nguyen-Phuc}, {Timmons}, {van Bezooijen}, {Weingrod},
  {Brookner}, {Butcher}, {Dougherty}, {Eder}, {Knagenhjelm}, {Larsen},
  {Mansir}, {Phan}, {Boyle}, {Cheimets}, {DeLuca}, {Golub}, {Gates}, {Hertz},
  {McKillop}, {Park}, {Perry}, {Podgorski}, {Reeves}, {Saar}, {Testa}, {Tian},
  {Weber}, {Dunn}, {Eccles}, {Jaeggli}, {Kankelborg}, {Mashburn}, {Pust},
  {Springer}, {Carvalho}, {Kleint}, {Marmie}, {Mazmanian}, {Pereira}, {Sawyer},
  {Strong}, {Worden}, {Carlsson}, {Hansteen}, {Leenaarts}, {Wiesmann},
  {Aloise}, {Chu}, {Bush}, {Scherrer}, {Brekke}, {Martinez-Sykora}, {Lites},
  {McIntosh}, {Uitenbroek}, {Okamoto}, {Gummin}, {Auker}, {Jerram}, {Pool}, \&
  {Waltham}}]{depont:2014}
{De Pontieu}, B., {Title}, A.~M., {Lemen}, J.~R., {et~al.} 2014, \solphys, 289,
  2733

\bibitem[{{Fedun} {et~al.}(2011{\natexlab{a}}){Fedun}, {Shelyag}, \&
  {Erd{\'e}lyi}}]{Fedun:2011c}
{Fedun}, V., {Shelyag}, S., \& {Erd{\'e}lyi}, R. 2011{\natexlab{a}}, \apj, 727,
  17

\bibitem[{{Fedun} {et~al.}(2011{\natexlab{b}}){Fedun}, {Verth}, {Jess}, \&
  {Erd{\'e}lyi}}]{Fedun:2011b}
{Fedun}, V., {Verth}, G., {Jess}, D.~B., \& {Erd{\'e}lyi}, R.
  2011{\natexlab{b}}, \apjl, 740, L46

\bibitem[{{Goossens} {et~al.}(2014){Goossens}, {Soler}, {Terradas}, {Van
  Doorsselaere}, \& {Verth}}]{Goos:2014}
{Goossens}, M., {Soler}, R., {Terradas}, J., {Van Doorsselaere}, T., \&
  {Verth}, G. 2014, \apj, 788, 9

\bibitem[{{Jess} {et~al.}(2009){Jess}, {Mathioudakis}, {Erd{\'e}lyi},
  {Crockett}, {Keenan}, \& {Christian}}]{Jess:2009}
{Jess}, D.~B., {Mathioudakis}, M., {Erd{\'e}lyi}, R., {et~al.} 2009, Science,
  323, 1582

\bibitem[{{Kitiashvili} {et~al.}(2013){Kitiashvili}, {Kosovichev}, {Lele},
  {Mansour}, \& {Wray}}]{Kitia:2013}
{Kitiashvili}, I.~N., {Kosovichev}, A.~G., {Lele}, S.~K., {Mansour}, N.~N., \&
  {Wray}, A.~A. 2013, \apj, 770, 37

\bibitem[{{Kitiashvili} {et~al.}(2012){Kitiashvili}, {Kosovichev}, {Mansour},
  \& {Wray}}]{Kitia:2012}
{Kitiashvili}, I.~N., {Kosovichev}, A.~G., {Mansour}, N.~N., \& {Wray}, A.~A.
  2012, \apjl, 751, L21

\bibitem[{{Kontogiannis} {et~al.}(2010{\natexlab{a}}){Kontogiannis},
  {Tsiropoula}, \& {Tziotziou}}]{Konto:2010a}
{Kontogiannis}, I., {Tsiropoula}, G., \& {Tziotziou}, K. 2010{\natexlab{a}},
  \aap, 510, A41

\bibitem[{{Kontogiannis} {et~al.}(2014){Kontogiannis}, {Tsiropoula}, \&
  {Tziotziou}}]{Konto:2014}
{Kontogiannis}, I., {Tsiropoula}, G., \& {Tziotziou}, K. 2014, \aap, 567, A62

\bibitem[{{Kontogiannis} {et~al.}(2010{\natexlab{b}}){Kontogiannis},
  {Tsiropoula}, {Tziotziou}, \& {Georgoulis}}]{Konto:2010b}
{Kontogiannis}, I., {Tsiropoula}, G., {Tziotziou}, K., \& {Georgoulis}, M.~K.
  2010{\natexlab{b}}, \aap, 524, A12

\bibitem[{{Leenaarts} {et~al.}(2009){Leenaarts}, {Carlsson}, {Hansteen}, \&
  {Rouppe van der Voort}}]{Leen:2009}
{Leenaarts}, J., {Carlsson}, M., {Hansteen}, V., \& {Rouppe van der Voort}, L.
  2009, \apjl, 694, L128

\bibitem[{{Leenaarts} {et~al.}(2012){Leenaarts}, {Carlsson}, \& {Rouppe van der
  Voort}}]{Leen:2012}
{Leenaarts}, J., {Carlsson}, M., \& {Rouppe van der Voort}, L. 2012, \apj, 749,
  136

\bibitem[{{Leenaarts} {et~al.}(2006){Leenaarts}, {Rutten}, {S{\"u}tterlin},
  {Carlsson}, \& {Uitenbroek}}]{Leen:2006}
{Leenaarts}, J., {Rutten}, R.~J., {S{\"u}tterlin}, P., {Carlsson}, M., \&
  {Uitenbroek}, H. 2006, \aap, 449, 1209

\bibitem[{{Lemen} {et~al.}(2012){Lemen}, {Title}, {Akin}, {Boerner}, {Chou},
  {Drake}, {Duncan}, {Edwards}, {Friedlaender}, {Heyman}, {Hurlburt}, {Katz},
  {Kushner}, {Levay}, {Lindgren}, {Mathur}, {McFeaters}, {Mitchell}, {Rehse},
  {Schrijver}, {Springer}, {Stern}, {Tarbell}, {Wuelser}, {Wolfson}, {Yanari},
  {Bookbinder}, {Cheimets}, {Caldwell}, {Deluca}, {Gates}, {Golub}, {Park},
  {Podgorski}, {Bush}, {Scherrer}, {Gummin}, {Smith}, {Auker}, {Jerram},
  {Pool}, {Soufli}, {Windt}, {Beardsley}, {Clapp}, {Lang}, \&
  {Waltham}}]{Lemen:2012}
{Lemen}, J.~R., {Title}, A.~M., {Akin}, D.~J., {et~al.} 2012, \solphys, 275, 17

\bibitem[{{Luna} {et~al.}(2015){Luna}, {Moreno-Insertis}, \&
  {Priest}}]{Luna:2015}
{Luna}, M., {Moreno-Insertis}, F., \& {Priest}, E. 2015, \apjl, 808, L23

\bibitem[{{Mart{\'{\i}}nez Pillet} {et~al.}(2011){Mart{\'{\i}}nez Pillet}, {Del
  Toro Iniesta}, {{\'A}lvarez-Herrero}, {Domingo}, {Bonet}, {Gonz{\'a}lez
  Fern{\'a}ndez}, {L{\'o}pez Jim{\'e}nez}, {Pastor}, {Gasent Blesa}, {Mellado},
  {Piqueras}, {Aparicio}, {Balaguer}, {Ballesteros}, {Belenguer}, {Bellot
  Rubio}, {Berkefeld}, {Collados}, {Deutsch}, {Feller}, {Girela}, {Grauf},
  {Heredero}, {Herranz}, {Jer{\'o}nimo}, {Laguna}, {Meller}, {Men{\'e}ndez},
  {Morales}, {Orozco Su{\'a}rez}, {Ramos}, {Reina}, {Ramos},
  {Rodr{\'{\i}}guez}, {S{\'a}nchez}, {Uribe-Patarroyo}, {Barthol}, {Gandorfer},
  {Knoelker}, {Schmidt}, {Solanki}, \& {Vargas Dom{\'{\i}}nguez}}]{Mart:2011}
{Mart{\'{\i}}nez Pillet}, V., {Del Toro Iniesta}, J.~C., {{\'A}lvarez-Herrero},
  A., {et~al.} 2011, \solphys, 268, 57

\bibitem[{{McAteer} {et~al.}(2003){McAteer}, {Gallagher}, {Williams},
  {Mathioudakis}, {Bloomfield}, {Phillips}, \& {Keenan}}]{McAteer:2003}
{McAteer}, R.~T.~J., {Gallagher}, P.~T., {Williams}, D.~R., {et~al.} 2003,
  \apj, 587, 806

\bibitem[{{McAteer} {et~al.}(2002){McAteer}, {Gallagher}, {Williams},
  {Mathioudakis}, {Phillips}, \& {Keenan}}]{McAteer:2002}
{McAteer}, R.~T.~J., {Gallagher}, P.~T., {Williams}, D.~R., {et~al.} 2002,
  \apjl, 567, L165

\bibitem[{{Michalitsanos}(1973)}]{Mich:1973}
{Michalitsanos}, A.~G. 1973, \solphys, 30, 47

\bibitem[{{Moll} {et~al.}(2011{\natexlab{a}}){Moll}, {Cameron}, \&
  {Sch{\"u}ssler}}]{Moll:2011b}
{Moll}, R., {Cameron}, R.~H., \& {Sch{\"u}ssler}, M. 2011{\natexlab{a}}, \aap,
  533, A126

\bibitem[{{Moll} {et~al.}(2011{\natexlab{b}}){Moll}, {Pietarila Graham},
  {Pratt}, {Cameron}, {M{\"u}ller}, \& {Sch{\"u}ssler}}]{Moll:2011}
{Moll}, R., {Pietarila Graham}, J., {Pratt}, J., {et~al.} 2011{\natexlab{b}},
  \apj, 736, 36

\bibitem[{{Park} {et~al.}(2016){Park}, {Tsiropoula}, {Kontogiannis},
  {Tziotziou}, {Scullion}, \& {Doyle}}]{Park:2016}
{Park}, S.-H., {Tsiropoula}, G., {Kontogiannis}, I., {et~al.} 2016, \aap, 586,
  A25

\bibitem[{{Pesnell} {et~al.}(2012){Pesnell}, {Thompson}, \&
  {Chamberlin}}]{Pesnell:2012}
{Pesnell}, W.~D., {Thompson}, B.~J., \& {Chamberlin}, P.~C. 2012, \solphys,
  275, 3

\bibitem[{{Scharmer} {et~al.}(2003){Scharmer}, {Bjelksjo}, {Korhonen},
  {Lindberg}, \& {Petterson}}]{Scharmer:2003a}
{Scharmer}, G.~B., {Bjelksjo}, K., {Korhonen}, T.~K., {Lindberg}, B., \&
  {Petterson}, B. 2003, in \procspie, Vol. 4853, Innovative Telescopes and
  Instrumentation for Solar Astrophysics, ed. S.~L. {Keil} \& S.~V. {Avakyan},
  341--350

\bibitem[{{Scharmer} {et~al.}(2008){Scharmer}, {Narayan}, {Hillberg}, {de la
  Cruz Rodriguez}, {L{\"o}fdahl}, {Kiselman}, {S{\"u}tterlin}, {van Noort}, \&
  {Lagg}}]{Scharmer:2008}
{Scharmer}, G.~B., {Narayan}, G., {Hillberg}, T., {et~al.} 2008, \apjl, 689,
  L69

\bibitem[{{Scherrer} {et~al.}(2012){Scherrer}, {Schou}, {Bush}, {Kosovichev},
  {Bogart}, {Hoeksema}, {Liu}, {Duvall}, {Zhao}, {Title}, {Schrijver},
  {Tarbell}, \& {Tomczyk}}]{Scher:2012}
{Scherrer}, P.~H., {Schou}, J., {Bush}, R.~I., {et~al.} 2012, \solphys, 275,
  207

\bibitem[{{Spruit}(1981)}]{Spruit:1981}
{Spruit}, H.~C. 1981, \aap, 98, 155

\bibitem[{{Stangalini} {et~al.}(2015){Stangalini}, {Giannattasio}, \&
  {Jafarzadeh}}]{Stanga:2015}
{Stangalini}, M., {Giannattasio}, F., \& {Jafarzadeh}, S. 2015, \aap, 577, A17

\bibitem[{{Stein} \& {Nordlund}(2000)}]{Stein:2000b}
{Stein}, R.~F. \& {Nordlund}, {\AA}. 2000, in Annals of the New York Academy of
  Sciences, Vol. 898, Astrophysical Turbulence and Convection, ed. J.~R.
  {Buchler} \& H.~{Kandrup}, 21

\bibitem[{{Stenflo}(1975)}]{Sten:1975}
{Stenflo}, J.~O. 1975, \solphys, 42, 79

\bibitem[{{Suematsu}(1990)}]{Suem:1990}
{Suematsu}, Y. 1990, in Lecture Notes in Physics, Berlin Springer Verlag, Vol.
  367, Progress of Seismology of the Sun and Stars, ed. Y.~{Osaki} \&
  H.~{Shibahashi}, 211

\bibitem[{{Torrence} \& {Compo}(1998)}]{Torr:1998}
{Torrence}, C. \& {Compo}, G.~P. 1998, Bulletin of the American Meteorological
  Society, 79, 61

\bibitem[{{Tsiropoula} {et~al.}(2012){Tsiropoula}, {Tziotziou}, {Kontogiannis},
  {Madjarska}, {Doyle}, \& {Suematsu}}]{Tsir:2012}
{Tsiropoula}, G., {Tziotziou}, K., {Kontogiannis}, I., {et~al.} 2012, \ssr,
  169, 181

\bibitem[{{Tziotziou} {et~al.}(2018){Tziotziou}, {Tsiropoula}, {Kontogiannis},
  {Scullion}, \& {Doyle}}]{tzio:2018}
{Tziotziou}, K., {Tsiropoula}, G., {Kontogiannis}, I., {Scullion}, E., \&
  {Doyle}, J.~G. 2018, \aap, 618, A51

\bibitem[{{Tziotziou} {et~al.}(2004){Tziotziou}, {Tsiropoula}, \&
  {Mein}}]{Tzio:2004}
{Tziotziou}, K., {Tsiropoula}, G., \& {Mein}, P. 2004, \aap, 423, 1133

\bibitem[{{Wedemeyer} \& {Steiner}(2014)}]{Wede:2014}
{Wedemeyer}, S. \& {Steiner}, O. 2014, \pasj, 66, S10

\bibitem[{{Wedemeyer-B{\"o}hm} \& {Rouppe van der Voort}(2009)}]{Wede:2009}
{Wedemeyer-B{\"o}hm}, S. \& {Rouppe van der Voort}, L. 2009, \aap, 507, L9

\bibitem[{{Wedemeyer-B{\"o}hm} {et~al.}(2012){Wedemeyer-B{\"o}hm}, {Scullion},
  {Steiner}, {Rouppe van der Voort}, {de La Cruz Rodriguez}, {Fedun}, \&
  {Erd{\'e}lyi}}]{Wede:2012}
{Wedemeyer-B{\"o}hm}, S., {Scullion}, E., {Steiner}, O., {et~al.} 2012, \nat,
  486, 505

\bibitem[{{Worrall}(2002)}]{Worr:2002}
{Worrall}, G. 2002, \mnras, 335, 628

\end{thebibliography}

\end{document}